\documentclass[10pt,twocolumn,letterpaper]{article}

\usepackage{cvpr}              %

\usepackage{graphicx}
\usepackage{amsmath}
\usepackage{amssymb}
\usepackage{booktabs}
\usepackage{float}
\usepackage{color}
\usepackage[english]{babel}
\usepackage{multirow}
\usepackage{makecell}
\usepackage{url}
\usepackage{dsfont}
\usepackage{array}
\usepackage{setspace}
\usepackage{arydshln}
\usepackage{bm}

\usepackage{array}
\usepackage{booktabs}

\usepackage{diagbox}

\usepackage{xcolor}

\usepackage[pagebackref,breaklinks,colorlinks]{hyperref}

\usepackage[capitalize]{cleveref}
\crefname{section}{Sec.}{Secs.}
\Crefname{section}{Section}{Sections}
\Crefname{table}{Table}{Tables}
\crefname{table}{Tab.}{Tabs.}

\def\cvprPaperID{0030} %
\def\confName{CVPR}
\def\confYear{2022}

\begin{document}

\title{NeRF-Editing: Geometry Editing of Neural Radiance Fields}

\author{\textbf{Yu-Jie Yuan} \textsuperscript{1,2,$\dagger$}
\quad
\textbf{Yang-Tian Sun} \textsuperscript{1,2,$\dagger$}
\quad
\textbf{Yu-Kun Lai} \textsuperscript{3}
\\
\textbf{Yuewen Ma} \textsuperscript{4}
\quad
\textbf{Rongfei Jia} \textsuperscript{4}
\quad
\textbf{Lin Gao} \textsuperscript{1,2*}
\\
\textsuperscript{1}Beijing Key Laboratory of Mobile Computing and Pervasive Device,\\ Institute of Computing Technology, Chinese Academy of Sciences 
\\
\textsuperscript{2}School of Computer and Control Engineering, University of Chinese Academy of Sciences \\
\textsuperscript{3}School of Computer Science \& Informatics, Cardiff University \quad
\textsuperscript{4}Alibaba Group \\
{\tt\small \{yuanyujie, sunyangtian, gaolin\}@ict.ac.cn} \quad
{\tt\small LaiY4@cardiff.ac.uk} \\
{\tt\small \{yuewen.my, rongfei.jrf\}@alibaba-inc.com}
}

\maketitle
\if TT\insert\footins{\noindent\footnotesize{
$\dagger$: Authors contributed equally \\
*Corresponding Author is Lin Gao (gaolin@ict.ac.cn)}}\fi

\begin{figure*}[t]
\begin{center}  
\includegraphics[width=0.9\linewidth]{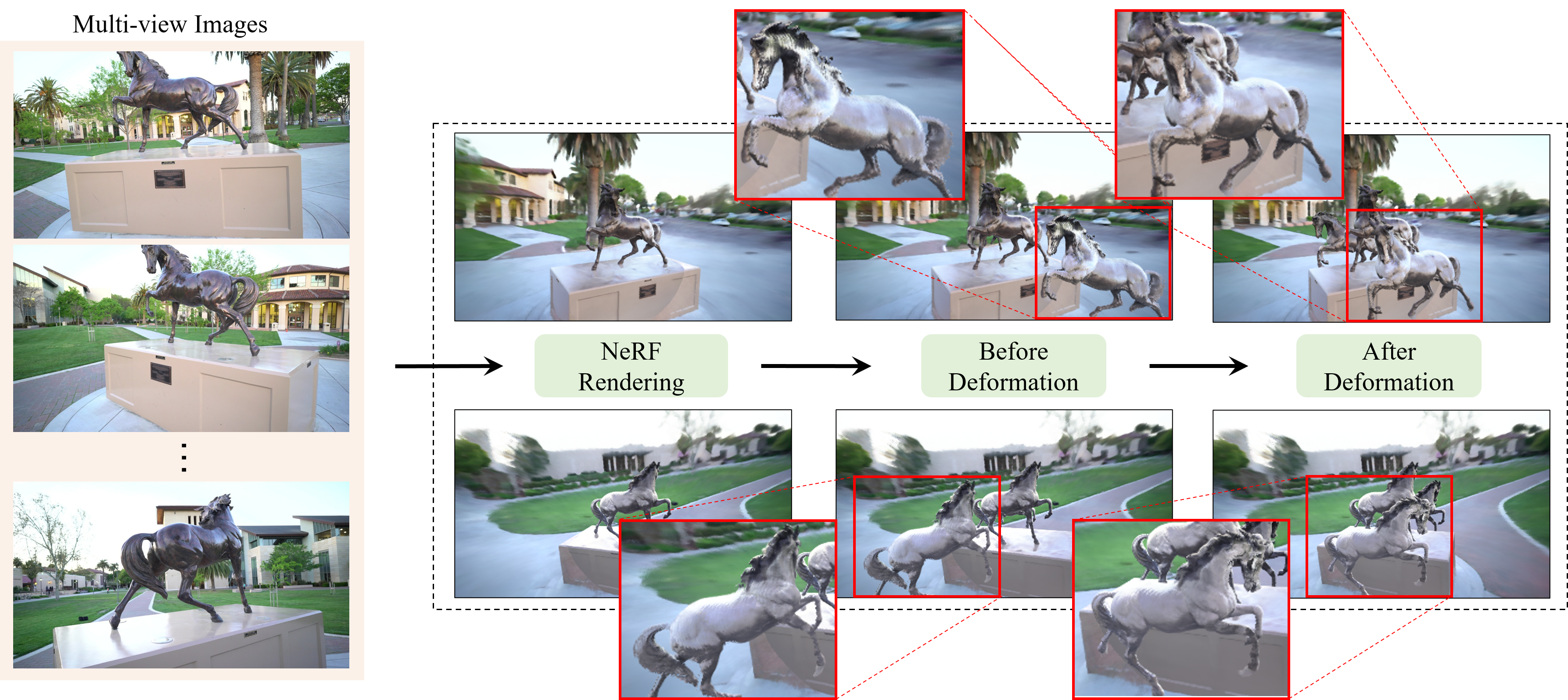}
\vspace{-6mm}
\end{center}
\caption{We propose a method to edit a static neural radiance field (NeRF). Users only need to capture multi-view images to build a NeRF representation, and then they can explicitly and intuitively edit the implicit representation of the scene. 
{Our method can perform user-controlled shape deformation on the geometry of the scene, which contains multiple objects.}
}
\label{fig:teaser}
\vspace{-6mm}
\end{figure*}

\begin{abstract}
 Implicit neural rendering, especially Neural Radiance Field (NeRF), has shown great potential in novel view synthesis of a scene. However, {current NeRF-based methods cannot enable users to perform user-controlled shape deformation in the scene.}
While existing works have proposed some approaches to modify the radiance field according to the user's constraints, the modification is limited to color editing or object translation and rotation.
 In this paper, we propose a method that allows users to 
 {perform controllable shape deformation on}
 the implicit representation of the scene, and synthesizes the novel view images of the edited scene without re-training the network. Specifically, we establish a correspondence between the extracted explicit mesh representation and the implicit neural representation of the target scene. Users can first utilize well-developed mesh-based deformation methods to deform the mesh representation of the scene. 
 {Our method then utilizes user edits from the mesh representation to bend the camera rays by introducing a tetrahedra mesh as a proxy, obtaining the rendering results of the edited scene.}
 Extensive experiments demonstrate that our framework can achieve ideal editing results not only on synthetic data, but also on real scenes captured by users.
\end{abstract}

\section{Introduction}
\label{sec:intro}
Novel view synthesis has been extensively studied in computer vision and computer graphics. In particular, the recently proposed neural radiance field (NeRF)~\cite{mildenhall2020nerf} has inspired a large number of follow-up works aiming to achieve better visual effects~\cite{liu2020neural}, faster rendering speed~\cite{yu2021plenoctrees,garbin2021fastnerf}, generalization to different scenes~\cite{yu2020pixelnerf}, relighting~\cite{boss2021nerd,srinivasan2021nerv}, applying to dynamic scenes~\cite{pumarola2021d}, and reducing the number of inputs~\cite{jain2021putting}. 
However, as an implicit modeling method, the neural radiance field is difficult for users to 
{edit}
or modify the scene objects, which is relatively easy with the explicit representation. The mesh representation, as a kind of explicit representation, is commonly used in shape modeling and rendering. There is a lot of research work on mesh deformation or editing~\cite{yuan2021revisit}. However, it is difficult to obtain an accurate explicit representation of a real-world scene. From a sparse set of images, one can use some Multi-View Stereo (MVS) method~\cite{schoenberger2016mvs} to reconstruct the point cloud or mesh representation of the scene, but the quality is generally poor. Rendering the reconstructed representation under novel views will lead to unrealistic results. Therefore, based on the promising novel view synthesis ability of implicit representations, such as NeRFs, further studying how to edit the implicit representation has become a new exploration direction.

Some works have already studied how to edit NeRF. For example, EditingNeRF~\cite{liu2021editing} was the first to propose editing on the implicit radiance field. They train on a set of synthetic models from the same category, such as chairs and tables from ShapeNet~\cite{chang2015shapenet}, and introduce shape code and color code to represent the geometry and appearance of different models, respectively. 
The user selects a desired color and draw a few coarse scribbles on an image of a specified view to indicate what should be changed. Then local edits are propagated to 3D regions through updating the network based on the loss between the original image and the edited image. This work is limited to color modification or the removal of certain parts of the shape, and it is impossible to make substantial modifications to the shape, {such as shape deformation}. A recent work, ObjectNeRF~\cite{yang2021objectnerf}, proposed to learn a decompositional neural radiance field, which separates the objects and the background. As such, it can duplicate, move or rotate the objects for editable scene rendering. However, it does not support {shape deformation} either. Meanwhile, some works~\cite{tretschk2021non,pumarola2021d} consider using NeRF to model dynamic scenes and using Multi-Layer Perceptron (MLP) to predict scene changes. However, they either limit the edits to human bodies~\cite{zhang2021editable,peng2021animatable}, or can only learn motion information from the recorded videos, and cannot perform active editing~\cite{pumarola2021d}.

In this paper, we propose a method for editing neural radiance field that combines the advantage of explicit representations for easy local editing and the advantage of implicit representations for realistic rendering effects. Different from the previous work~\cite{liu2021editing,yang2021objectnerf}, we focus on the geometric content of the scene, {as shown in Fig.~\ref{fig:teaser},} supporting users to edit the scene geometry, and can perform photo-realistic rendering from novel views. As far as we know, 
{we are the first to perform user-controlled shape deformation on the NeRF of general scenes.} To this end, we first extract an explicit triangular mesh representation from the trained NeRF. The explicit mesh representation is then intuitively deformed by the user. Next, a tetrahedral mesh is built from the triangular mesh representation, which wraps around the triangular mesh. We use the deformation of the triangular mesh to drive the deformation of the tetrahedral mesh, which propagates the deformation of the scene geometric surface to the spatial discrete deformation field. Finally, we use tetrahedral vertex interpolation to complete the propagation from the discrete deformation field to the continuous deformation field. The rays passing through the tetrahedral mesh will be bent accordingly following the continuous deformation field, so that the final rendering result conforms to the user's edits. Our method is general, not limited to specific shapes such as human bodies, and applicable to arbitrary shapes such as animal models and general man-made objects.

\section{Related Work}
Our NeRF editing framework provides a new paradigm for novel view synthesis of an edited neural implicit scene representation. Here, we summarize related work of novel view synthesis and 3D deformation/editing methods.

\textbf{Novel view synthesis.}
To infer the photo-realistic novel view synthesis result from given input images, prior works rely on explicit~\cite{Chaurasia2013DepthSA,Hedman2017Casual3P,Snavely2006PhotoTE,Hedman2016ScalableII} or implicit~\cite{Gortler1996TheL,Levoy1996LightFR,szeliski1998stereo} geometry representation of the real world scene. Recently, {%
used both as a component in deep neural network pipelines and as a standalone rendering pipeline, neural rendering has achieved immense progress, which is comprehensively summarized in ~\cite{Tewari2020StateOT, Tewari2021AdvancesIN}}. It adopts deep neural networks to synthesize images, which can be employed on multiple representations, such as voxels~\cite{sitzmann2019deepvoxels,lombardi2019neural}, point clouds~\cite{aliev2020neural,dai2020neural}, meshes~\cite{chen2018deep,thies2019deferred,riegler2020free,Riegler2021SVS}, multi-plane images (MPIs)~\cite{zhou2018stereo,mildenhall2019local,li2020crowdsampling} and implicit fields~\cite{sitzmann2019scene,kellnhofer2021neural}. As one of the representative works, Neural Radiance Field (NeRF)~\cite{mildenhall2020nerf} has attracted a lot of attention, which uses a multi-layer perceptron (MLP) to model the geometry and appearance of a scene. NeRF can achieve  photo-realistic synthesis of novel view images with view-dependent effects. However, NeRF still has shortcomings and plenty of work has extended the original NeRF, including better synthesis effects~\cite{liu2020neural,wizadwongsa2021nex,Zhang2020NeRFAA}, applicable to dynamic scenes~\cite{pumarola2021d,tretschk2021non,peng2021animatable,peng2021neural,gafni2021dynamic,zhang2021editable,xian2021space,li2021nsff,park2021nerfies,park2021hypernerf}, faster rendering speed~\cite{garbin2021fastnerf,hedman2021baking,reiser2021kilonerf,yu2021plenoctrees}, generalization to different scenes~\cite{wang2021ibrnet,chen2021mvsnerf}, relighting~\cite{srinivasan2021nerv,boss2021nerd,zhang2021nerfactor}, etc. NeRF-related works have been summarized in \cite{dellaert2021neural}. In this work, we focus on geometry editing/deformation for NeRF. As mentioned before, EditingNeRF~\cite{liu2021editing} proposes to edit on the rendered image and uses network optimization to achieve editing transfer to the entire image and novel view images. However, the edits are limited to 2D images, which cannot change the spatial position of the object, let alone change the shape of the object. ObjectNeRF~\cite{yang2021objectnerf} has a decompositional network architecture, which can only duplicate, move or rotate objects. Our framework, however, support editing the geometric shape of the objects in NeRF, which can then be used to synthesize photo-realistic novel view images for visualization.

\textbf{3D deformation and editing methods.}
Editing a 3D model means deforming the shape of the model under some controls given by the user.
There has been much work about the editing of explicit geometry representation~\cite{gao2012p,chen2017rigidity}, which we refer readers to a recent survey~\cite{yuan2021revisit}. Traditional mesh deformation methods are based on Laplacian coordinates~\cite{sorkine2004laplacian,sorkine2005laplacian,lipman2005laplacian}, Poisson equation~\cite{yu2004mesh}, and dual Laplacian coordinates~\cite{au2006dual}. As a representative work among them, ARAP (As-Rigid-As-Possible) deformation~\cite{Sorkine2007arap} is an interactive mesh editing scheme, which preserves details during the deformation by maintaining the rigidity of the local transformations. Another approach to driving mesh deformation is through a proxy, such as skeletons~\cite{magnenat1988joint,jacobson2012fast} or cages~\cite{sederberg1986free,zhang2020proxy,wang2020NeuralCage}. These methods need to calculate the weights~\cite{jacobson2011bounded,yuan2019data,floater2003mean} between the proxy and the mesh vertices, and propagate the transformation of the proxy to the mesh. With the proliferation of geometric models~\cite{Bogo2014faust}, data-driven deformation~\cite{sumner2005mesh,gao2016efficient,gao2019sparse} becomes available which analyzes the deformation prior of existing shapes in the dataset and produces more realistic results. At the same time, plenty of data also allows neural networks to be introduced into 3D editing~\cite{tan2018variational,yang2021multiscale,neuroskinning2019,RigNet}. In addition to the explicit mesh representation, the implicit field can also be edited in combination with a neural network. Deng~\etal proposed the deformed implicit field~\cite{deng2021deformed}, which is capable of modeling dense surface correspondence and shape editing based on the learned information from an object category. Our work also aims to edit implicit representations, in particular NeRFs. The difference is that we take advantage of the intuitive and convenient characteristics of explicit mesh editing. By establishing a connection between the explicit mesh representation and implicit neural representation, well-developed mesh deformation methods are used to edit the geometry of implicit representation.

\section{Our Method}
\begin{figure*}
    \centering
    \includegraphics[width=1.0\linewidth]{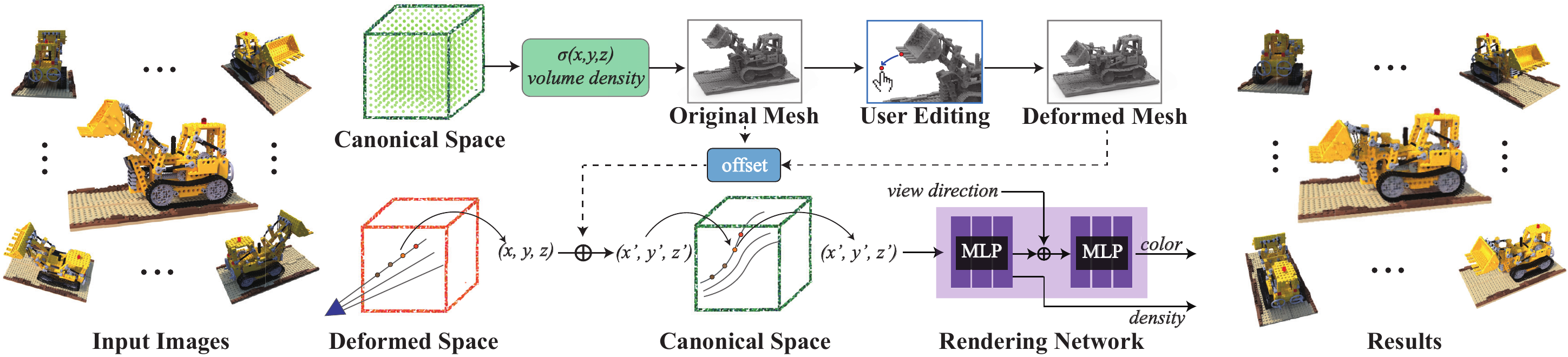}
    \vspace{-6mm}
    \caption{The pipeline of our NeRF editing framework. The user edits the reconstructed mesh and a continuous deformation field is built to bend the rays accordingly.}
    \vspace{-6mm}
    \label{fig:pipeline}
\end{figure*}

Our work is based on the neural radiance field (NeRF)~\cite{mildenhall2020nerf}, which has promising performance in novel view synthesis. As a result, our method enables users to 
{perform shape deformation on}
the content of the scene, and can generate new images from arbitrary views after editing. We will first briefly review NeRF pipeline (Sec.~\ref{sec:nerf}), and then introduce how to extract the explicit triangular mesh representation from the implicit representation of the scene and enable users to edit the mesh representation (Sec.~\ref{sec:mesh}). 
{After the user edits the triangular mesh representation of the scene, we need to transfer this deformation to the implicit volume representation. We split the transfer into two steps. The first step is to transfer the surface mesh deformation to a volumetric mesh, where we build a tetrahedra mesh that wraps around the surface mesh, and transfer user edits on the surface mesh to discrete deformation fields on the tetrahedra mesh (Sec.~\ref{sec:volume_deform}).
The next step is to transform the discrete deformation field to a continuous deformation field in the space volume, which is used to guide the bending of the rays to render images conforming to user edits (Sec.~\ref{sec:ray_blending}).
We will later show that directly transferring the deformation from surface mesh to the implicit volume by interpolation will lead to obvious artifacts compared to our two step strategy in Sec.~\ref{sec:compare}.}
Our method establishes the connection between the explicit mesh representation and the implicit radiance field, enabling users to modify the geometry of radiance field through intuitive edits. The pipeline is shown in Fig.~\ref{fig:pipeline}.

\subsection{Neural Radiance Fields} 
\label{sec:nerf}
Neural Radiance Field or NeRF~\cite{mildenhall2020nerf} proposes to use a multi-layer perceptron (MLP) network to model the geometry and appearance of the scene from a sparse set of images. 
{Given the known camera parameters,} the image pixels can be transformed to the world coordinate system and connected with the camera position to generate the light rays that are directed toward the scene. NeRF samples points on the ray and uses volume rendering~\cite{kajiya1984ray} to obtain the color of each ray. The spatial coordinates $\mathbf{p}=(x,y,z)$ of each sampled point and the ray direction $\mathbf{d}=(\theta,\phi)$ will go through positional encoding $\zeta(\cdot)$, and then input into the fully connected network to predict the volume density $\sigma$ and RGB value $\mathbf{c}$: $F_{\Theta}: (\zeta(\mathbf{p}),\zeta(\mathbf{d})) \rightarrow (\sigma, \mathbf{c})$,
where $\Theta$ represents the network weights. 
The predicted density value $\sigma$ can be interpreted as the differentiable probability of the ray terminated at the sampled point, and the color $\hat{C}(\mathbf{r})$ of the image pixel corresponding to the ray $\mathbf{r}(t)$ can be calculated through discrete integration:
\begin{align}
    \hat{C}(\mathbf{r}) = \sum^N_{i=1} {\exp(-\sum^{i-1}_{j=1}\sigma_j\delta_j)(1-\exp(-\sigma_i\delta_i))\mathbf{c}_i}, 
\label{equ:quadrature}
\end{align}
where $\delta_i=t_{i+1}-t_i$ is the distance between adjacent samples. The network is supervised by the RGB loss function, which is 
{calculated between the generated color $\hat{C}(\mathbf{r})$ and the ground truth color $C(\mathbf{r})$ of the ray.}

\subsection{Editing of Explicit Surface Mesh Representation}
\label{sec:mesh}
After the NeRF network is trained, an explicit triangular mesh representation can be extracted directly from the neural radiance fields using Marching Cubes~\cite{lorensen1987marching}. However, the mesh extracted from the original NeRF network is often with rough surface. 
In order to obtain a satisfactory 
{editing}
representation, we adopt the reconstruction method proposed in NeuS~\cite{Wang2021NeuS}, {which takes a bias-free volume rendering manner to learn the geometry as a neural signed distance function (SDF) representation.} The mesh representation {extracted from the zero-level set of SDF} will serve as the user's editing object, allowing users to edit the scene content intuitively. In this paper, we use the classic ARAP (as-rigid-as-possible) deformation method~\cite{Sorkine2007arap} to enable users to interactively deform the mesh. It should be noted that any other mesh deformation method can be used here, including skeleton-based and cage-based methods. 

The extracted triangular mesh is denoted as $S$, and $N(i)$ represents the index set of vertices adjacent to vertex $i$. We further denote $\bm{v}_i \in \mathbb{R}^3$ as the position of the vertex $i$ on the mesh $S$. After the user's edits, the mesh $S$ is transformed to the deformed mesh $S'$ with the same connectivity and different vertex positions $\bm{v}_i'$, treating user editing as boundary conditions. 
The overall ARAP deformation energy is to measure the rigidity of the entire mesh and is the sum of the distortion energies of each deformation cell, including vertex $i$ and its 1-ring neighbors, shown in Eq.~\ref{equ:arap}.
\begin{equation}
E(S')=\sum_{i=1}^{n} \sum_{j \in N(i)} w_{ij} {\| (\bm{v}'_i-\bm{v}'_j) - \mathbf{R}_i(\bm{v}_i-\bm{v}_j)\|}^2.
\label{equ:arap}
\end{equation}
Here, $w_{ij}= \frac{1}{2} (\cot{\alpha_{ij}} + \cot{\beta_{ij}})$ is the cotangent weight, and $\alpha_{ij}$, $\beta_{ij}$ are the angles opposite to the mesh edge $(i,j)$. 
$\mathbf{R}_i$ is the local rotation at vertex $i$. The deformed shape $S'$ is obtained by minimizing the ARAP energy, which can be efficiently solved by alternately optimizing local rotations $\mathbf{R}_i$ and deformed positions $\bm{v}'_i$. We refer the readers to \cite{Sorkine2007arap} for the specific optimization process.

\begin{figure}
    \centering
    \includegraphics[width=0.9\linewidth]{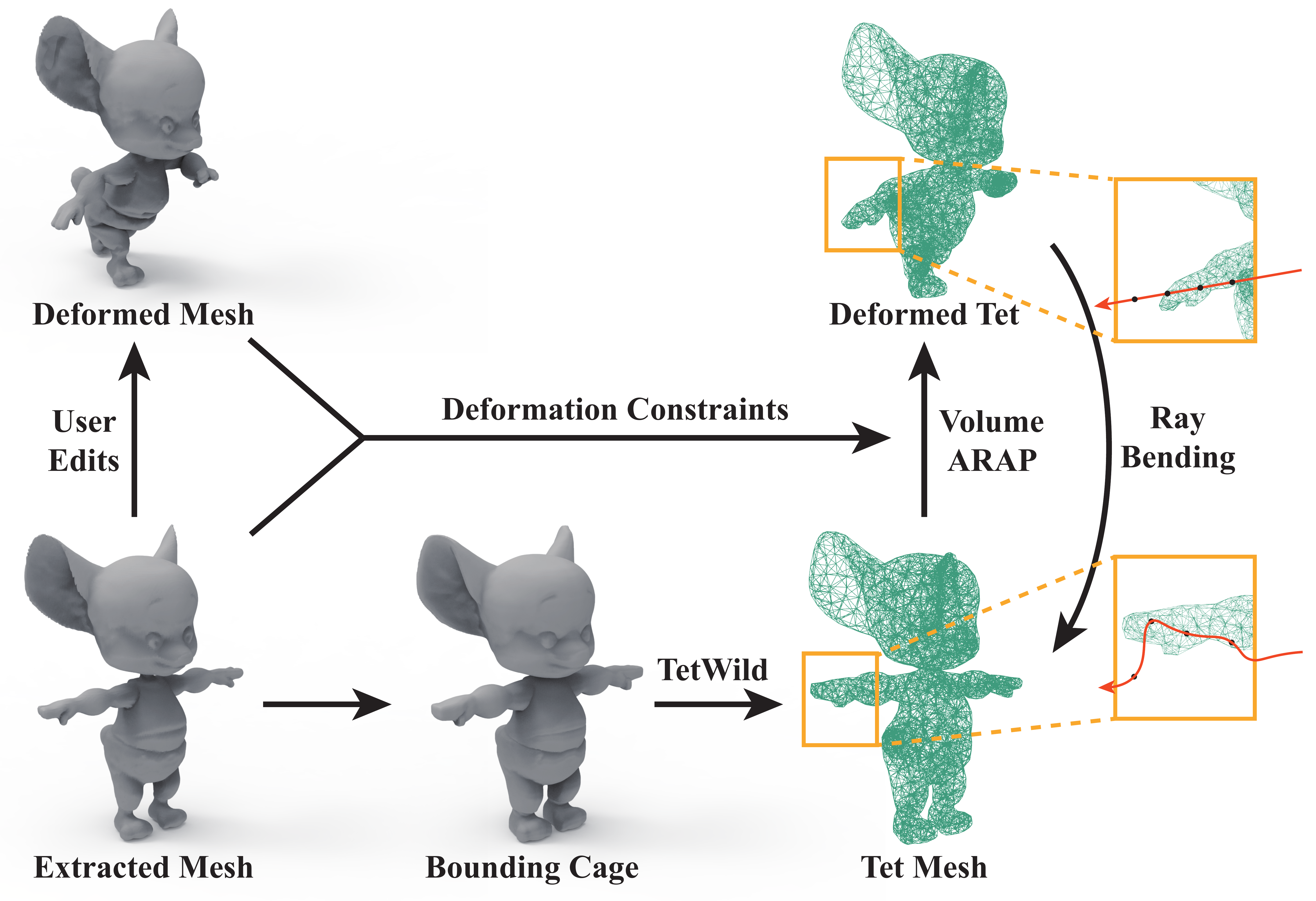}
    \vspace{-3mm}
    \caption{We use discrete deformations specified by users to bend the rays.}
    \label{fig:deformation}
    \vspace{-6mm}
\end{figure}

\subsection{Deformation Transfer to Discrete Volume}
\label{sec:volume_deform}
After the user edits the triangular mesh representation of the scene, the deformation needs to be transferred to the implicit volume representation. As introduced before, we split the transfer into two steps. In the first step, we build a tetrahedral mesh (a discrete volumetric representation) to cover the extracted triangular mesh. Starting from the extracted triangular mesh $S$, we first calculate a cage mesh that wraps the mesh $S$. This can be achieved by enlarging the triangular mesh by making a certain distance offset from the mesh surface in the normal direction. 
We set the default value to 5\% of the averaged distance from the camera position to object center.
The internal space of the cage mesh can be regarded as the {``effective space''} of the implicit volume, {because the area near real geometry surface of the scene is}
enclosed by this cage mesh. When editing larger scenes with multiple objects, this design also ensures other objects not being edited are not affected.
{We use the tetrahedralization method, TetWild~\cite{Hu2020FastTM}, to tetrahedronize the cage mesh to obtain a tetrahedral mesh representation $T$. It should be noted that the extracted triangular mesh $S$ is also wrapped in the tetrahedral mesh $T$. We visualize some extracted triangular mesh $S$ and the corresponding tetrahedral mesh $T$ in the supplementary material.} 
We use the displacement of the triangular mesh vertices $\bm{v}_i$ to drive the deformation of the tetrahedral mesh $T$, which transfers the surface deformation to the tetrahedral mesh. The deformed tetrahedral mesh is denoted as $T'$, and $\bm{t}_k$ and $\bm{t}'_k$ denote the vertices of the tetrahedral mesh before and after deformation respectively, {where $k$ is the vertex index}. Here, we also use the ARAP deformation method to deform the tetrahedral mesh $T$ under the constraints of the surface mesh deformation. Eq.~\ref{equ:arap} can be extended from the triangular mesh to the tetrahedral mesh straightforwardly. The only difference is that the constraints are changed from user-specified control points to the triangular mesh vertices. We can find which tetrahedron each triangular mesh vertex is located in, and calculate its barycentric coordinates relative to the four vertices of the tetrahedron. Then, the optimization problem is,
\begin{equation}
\min E(T') \text{, subject to } \bm{A} \bm{t}' = \bm{v}',
\label{equ:volume_arap}
\end{equation}
where $\bm{A}$ is the barycentric weight matrix. This optimization problem can be converted into linear equations using the Lagrangian multiplier method. Please refer to the supplementary material for the specific derivation.

\subsection{Ray Bending}
\label{sec:ray_blending}
After transferring the surface deformation to the tetrahedral mesh, we can obtain the discrete deformation field of the ``effective space''. We now utilize these discrete transformations to bend the casting rays. To generate an image of the deformed radiance field, we cast rays to the space containing the deformed tetrahedral mesh. For each sampled point on the ray, we find which tetrahedron of the deformed tetrahedral mesh $T'$ it is located in. Using the correspondence between $T$ and $T'$, the displacement from the vertices after deformation to the vertices before deformation can be obtained. Through barycentric interpolation of the displacements of the four vertices of the tetrahedron where the sampled point is located, the displacement of the sampled point back to the original ``effective space'' $\Delta p$ can be obtained. We add the displacement $\Delta p$ to the input coordinate of the sampled point to predict the density and RGB values.
\begin{equation}
    (\zeta(\mathbf{p}+\Delta \mathbf{p}), \zeta(\mathbf{d})) \rightarrow (\sigma, \mathbf{c}).
    \label{equ:deformed}
\end{equation}
The density and RGB values of the sampled points along the ray are used to calculate the corresponding pixel color using Eq.~\ref{equ:quadrature}. It should be noted that the sampled points that are not within the tetrahedral mesh $T'$ will not be moved, i.e., the part of the ray outside the tetrahedral mesh will not be bent. The process of building deformation field is illustrated in Fig.~\ref{fig:deformation}.

\begin{figure}[htb]
    \centering
    \setlength{\fboxrule}{0.5pt}
    \setlength{\fboxsep}{-0.1cm}
    \begin{spacing}{1}
    \renewcommand\arraystretch{-0.}
    \setlength{\tabcolsep}{-1pt}
    \begin{tabular}{p{0.07\linewidth}<{\centering}p{0.24\linewidth}<{\centering}p{0.24\linewidth}<{\centering}p{0.24\linewidth}<{\centering}p{0.24\linewidth}<{\centering}}
    
    & View1 &  View2 & View3 &  View4  \\
    
    \vspace{-15mm} \rotatebox{90}{Before} &
    \includegraphics[width=1\linewidth]{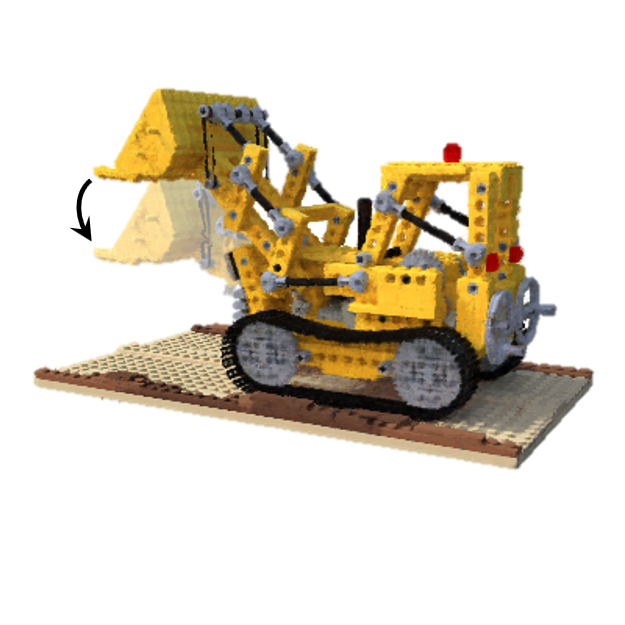} &
    \includegraphics[width=1\linewidth]{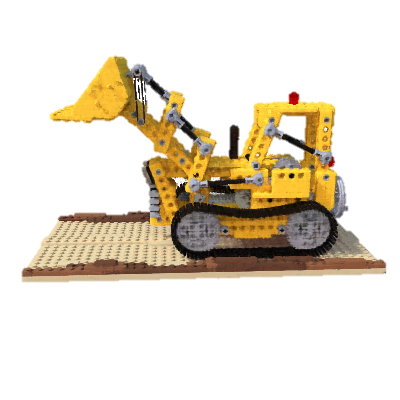} &
    \includegraphics[width=1\linewidth]{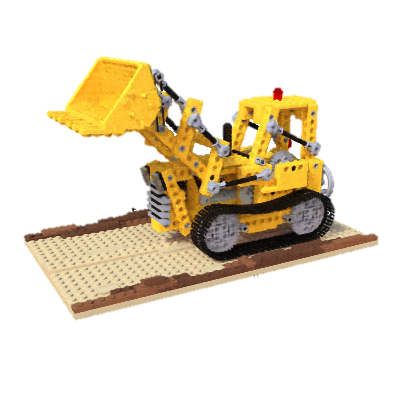} &
    \includegraphics[width=1\linewidth]{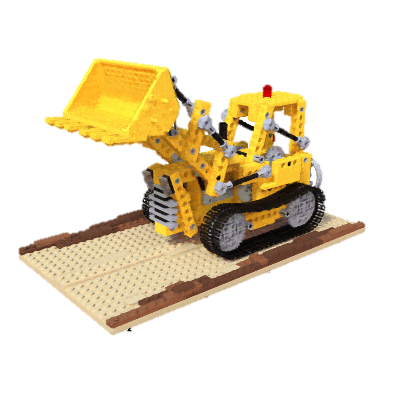}
    \\
    \specialrule{0em}{-6pt}{-6pt}
    \vspace{-15mm} \rotatebox{90}{After} &
    \includegraphics[width=1\linewidth]{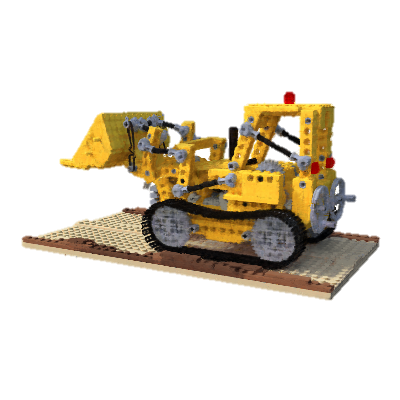} &
    \includegraphics[width=1\linewidth]{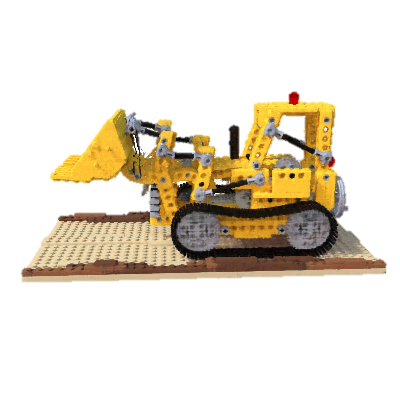} &
    \includegraphics[width=1\linewidth]{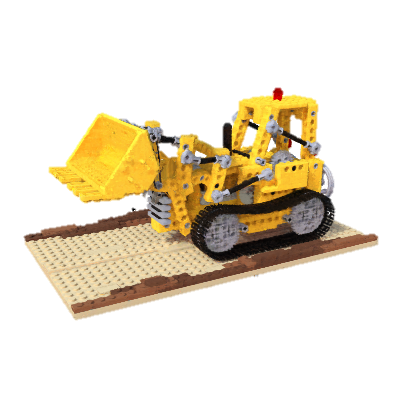} &
    \includegraphics[width=1\linewidth]{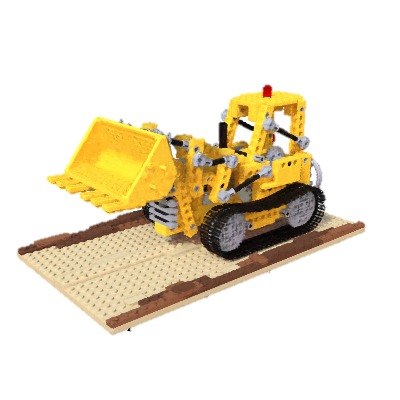}
    \\
   \specialrule{0em}{-5pt}{-5pt}
    \vspace{-15mm} \rotatebox{90}{Before} &
    \includegraphics[width=1.15\linewidth]{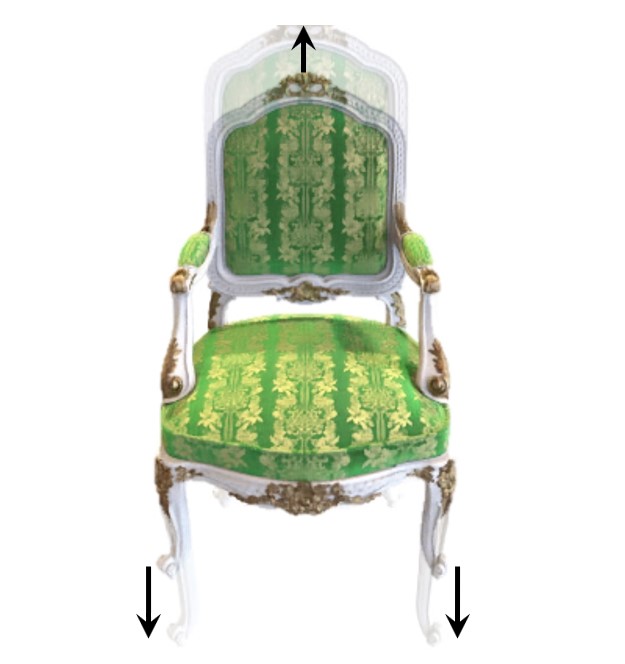} &
    \includegraphics[width=1.15\linewidth]{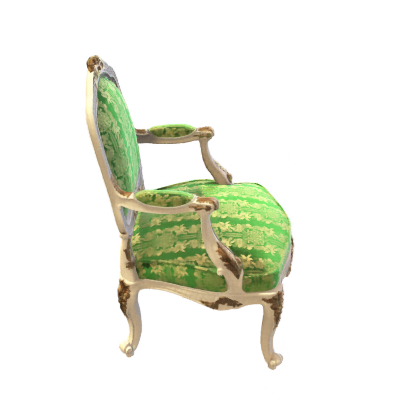} &
    \includegraphics[width=1.15\linewidth]{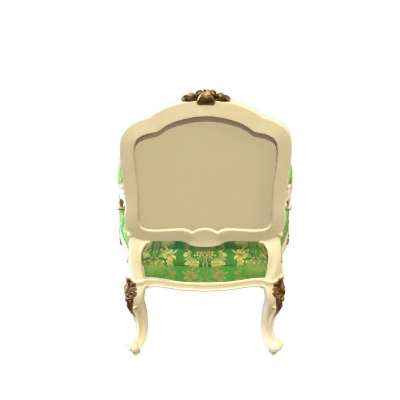} &
    \includegraphics[width=1.15\linewidth]{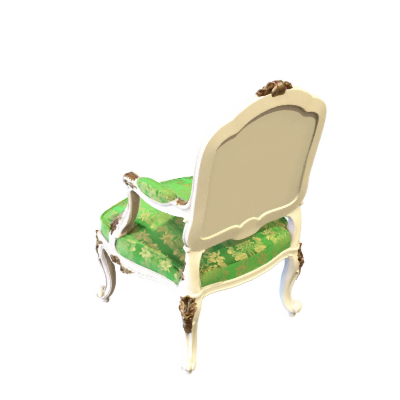}
    \\
    \specialrule{0em}{-1pt}{-1pt}
    \vspace{-15mm} \rotatebox{90}{After} &
    \includegraphics[width=1.15\linewidth]{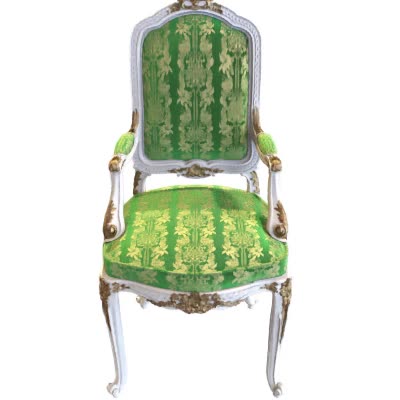} &
    \includegraphics[width=1.15\linewidth]{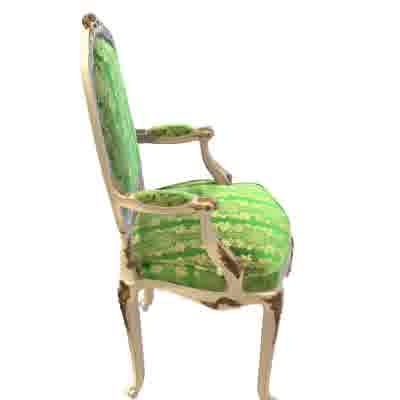} &
    \includegraphics[width=1.15\linewidth]{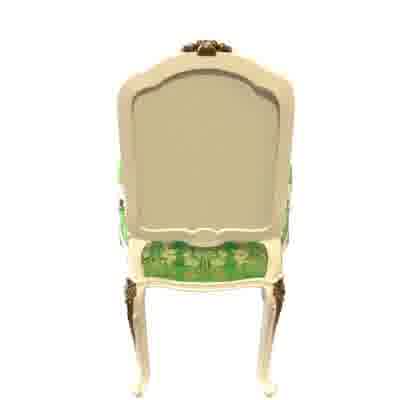} &
    \includegraphics[width=1.15\linewidth]{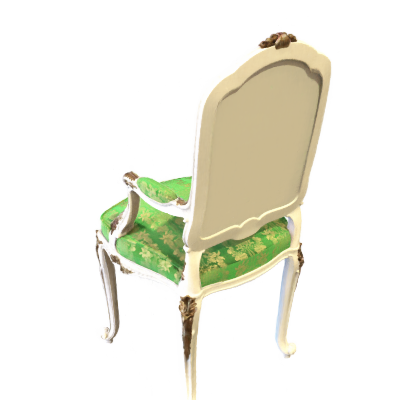}

    \end{tabular}
    \end{spacing}
    \vspace{-3mm}
    \caption{We show the editing results (row ``After'') compared with NeRF rendering results (row ``Before'') on synthetic data under different views, including a Lego bulldozer and a chair, which illustrate that our method can edit the NeRF of general models. Different columns show different views.}
    \vspace{-4mm}
    \label{fig:editing_results}
\end{figure}

\begin{figure}[htb]
    \centering
    \setlength{\fboxrule}{0.5pt}
    \setlength{\fboxsep}{-0.01cm}
    \begin{spacing}{1}
    \renewcommand\arraystretch{-0.}
    \setlength{\tabcolsep}{-0.8pt}
    \begin{tabular}{p{0.07\linewidth}<{\centering}p{0.19\linewidth}<{\centering}p{0.19\linewidth}<{\centering}p{0.19\linewidth}<{\centering}p{0.19\linewidth}<{\centering}p{0.19\linewidth}<{\centering}}

     & & View1 & View2 & View3 &  View4  \\
     \specialrule{0em}{2pt}{1pt}
    
    &
    \vspace{-10mm} \hspace{-5mm} \diagbox[innerleftsep=-10pt, innerrightsep=2pt]{Edit Op.}{Before \\ \vspace{1em}} &
    \includegraphics[width=1.15\linewidth,trim=35 0 15 30,clip]{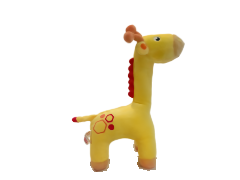} & %
    \includegraphics[width=1.15\linewidth,trim=35 0 15 30,clip]{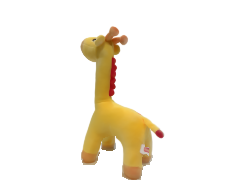} &
    \includegraphics[width=1.15\linewidth,trim=35 0 15 30,clip]{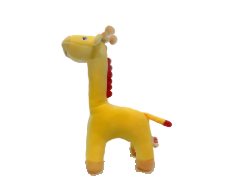} &
    \includegraphics[width=1.15\linewidth,trim=35 0 15 30,clip]{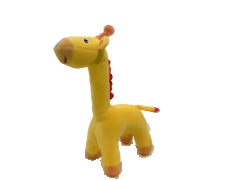}
    
    \\
    \specialrule{0em}{-12pt}{0pt}
    \vspace{-10mm} \rotatebox{90}{Edit1} &
    \includegraphics[width=1.15\linewidth,trim=20 0 10 20,clip]{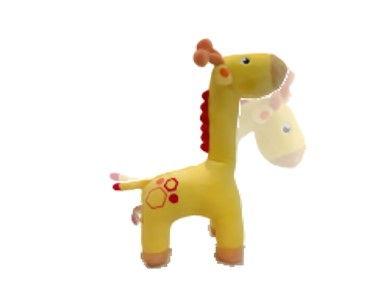} &
    \includegraphics[width=1.15\linewidth,trim=35 0 15 40,clip]{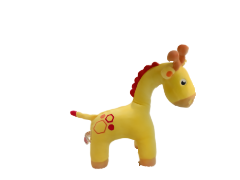} &
    \includegraphics[width=1.15\linewidth,trim=35 0 15 40,clip]{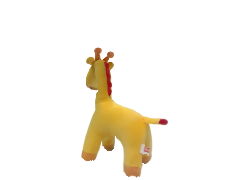} &
    \includegraphics[width=1.15\linewidth,trim=35 0 15 40,clip]{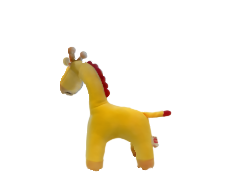} &
    \includegraphics[width=1.15\linewidth,trim=35 0 15 40,clip]{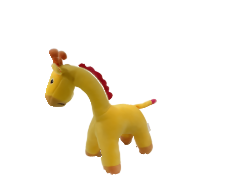}
    
    \\
    \specialrule{0em}{-7pt}{-5pt}
    \vspace{-10mm} \rotatebox{90}{Edit2} &
    \includegraphics[width=1.15\linewidth,trim=20 0 10 20,clip]{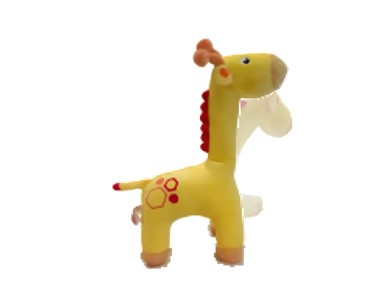} &
    \includegraphics[width=1.15\linewidth,trim=35 0 15 40,clip]{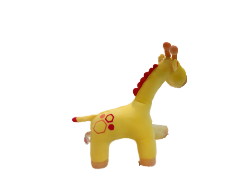} &
    \includegraphics[width=1.15\linewidth,trim=35 0 15 40,clip]{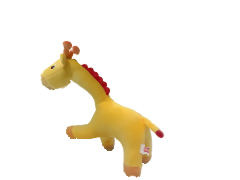} &
    \includegraphics[width=1.15\linewidth,trim=35 0 15 40,clip]{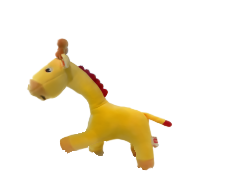} &
    \includegraphics[width=1.15\linewidth,trim=35 0 15 40,clip]{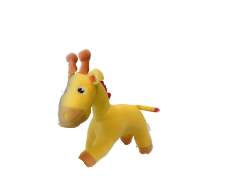}
    
    \\
    \specialrule{0em}{-5pt}{-5pt}
    \vspace{-10mm} \rotatebox{90}{Edit3} &
    \includegraphics[width=1.15\linewidth,trim=20 0 15 0,clip]{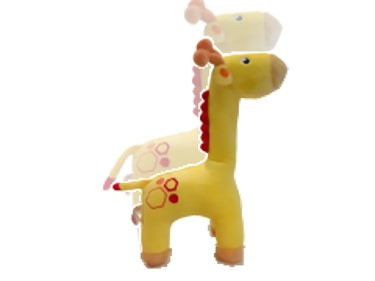} &
    \includegraphics[width=1.15\linewidth,trim=35 0 15 0,clip]{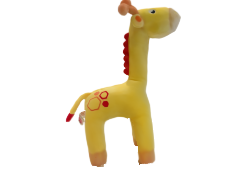} &
    \includegraphics[width=1.15\linewidth,trim=35 0 15 0,clip]{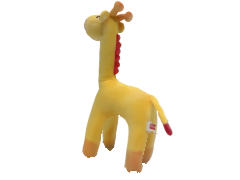} &
    \includegraphics[width=1.15\linewidth,trim=35 0 15 0,clip]{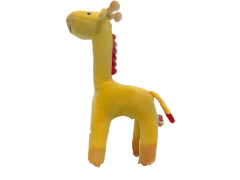} &
    \includegraphics[width=1.15\linewidth,trim=35 0 15 0,clip]{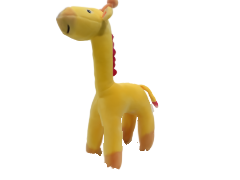}

    \end{tabular}
    \end{spacing}
    \vspace{-3mm}
    \caption{Results of our NeRF editing framework (rows ``Edit1'', ``Edit2'', ``Edit3'') compared with original NeRF results (row ``Before'') on a real captured giraffe soft toy. Different columns show different views. We can edit the object into different poses and render it under different views. ``Edit Op.'' denotes `` Edit Operation''.}
    \vspace{-4mm}
    \label{fig:editing_results_giraffe}
\end{figure}

\section{Experiments and Evaluations}
In this section, we conduct several qualitative and quantitative experiments, including showing editing results on both synthetic data and captured real scenes, comparisons with baseline methods, and ablation study.

\subsection{Datasets and metrics}
We demonstrate our method on several public synthetic data, including some characters in the mixamo~\cite{mixamo}, the Lego bulldozer and chair from NeRF~\cite{mildenhall2020nerf}.
Moreover, we also test our method on 
a real captured horse statue from FVS dataset~\cite{riegler2020free} and
several real scenes captured by ourselves. The characters in the mixamo are rendered by ourselves. We generate 100 random views from the upper hemisphere with Blender for training. For the data from NeRF datasets, we use the default training setting of the datasets. For the real scenes captured by ourselves, we leave one image for validation, and the other images are used for training. More information about self-captured dataset is included in supplementary document.

It needs to be noted that different from dynamic NeRF methods~\cite{pumarola2021d}, it is difficult to obtain the ground truths of the novel view synthesis results after user editing, especially on real scenes, as such edited scenes do not physically exist. So we mainly evaluate our approach quantitatively and qualitatively on the characters in the mixamo. Specifically, we rig the mixamo character model, render the deformed characters as the ground truths and compare them with the outputs of our NeRF editing method. We use Structural Similarity Index Measure (SSIM)~\cite{SSIM}, Learned Perceptual Image Patch Similarity (LPIPS)~\cite{zhang2018unreasonable} and Peak Signal-to-Noise Ratio (PSNR) as the metrics to evaluate the performance of our approach.
We also evaluate the Fréchet Inception Distance~\cite{Heusel2017GANsTB} (FID) score on real scenes to measure the similarity between the results before editing and after editing, since the ground truths are not indispensable.

\subsection{Editing Results}
\begin{figure}[thb]
    \centering
    \setlength{\fboxrule}{0.5pt}
    \setlength{\fboxsep}{-0.01cm}
    \begin{spacing}{1}
    \renewcommand\arraystretch{-0.}
    \setlength{\tabcolsep}{-0.8pt}
    \begin{tabular}{p{0.07\linewidth}<{\centering}p{0.24\linewidth}<{\centering}p{0.24\linewidth}<{\centering}p{0.24\linewidth}<{\centering}p{0.24\linewidth}<{\centering}}

     & View1 & \hspace{1mm} View2 & View3 &  View4  \\
    \vspace{-10mm} \rotatebox{90}{Before} &
    \includegraphics[width=1.15\linewidth]{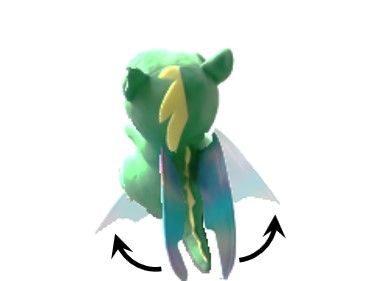} &
    \includegraphics[width=1.15\linewidth]{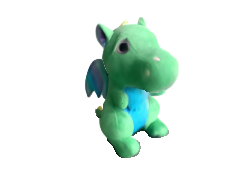} &
    \includegraphics[width=1.15\linewidth]{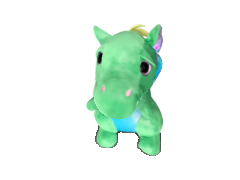} &
    \includegraphics[width=1.15\linewidth]{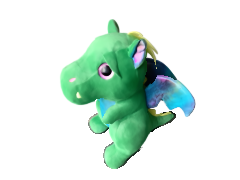}
    \\
    \specialrule{0em}{0pt}{0pt}
    \vspace{-10mm} \rotatebox{90}{After} &
    \includegraphics[width=1.15\linewidth]{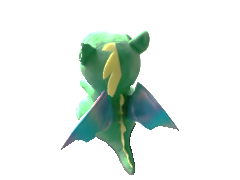} &
    \includegraphics[width=1.15\linewidth]{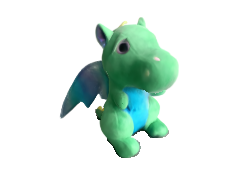} &
    \includegraphics[width=1.15\linewidth]{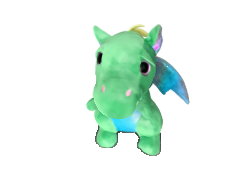} &
    \includegraphics[width=1.15\linewidth]{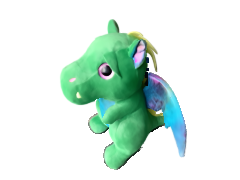}
    \\
    \specialrule{0em}{0pt}{0pt}
    \vspace{-12mm} \rotatebox{90}{Before} &
    \includegraphics[width=1.35\linewidth]{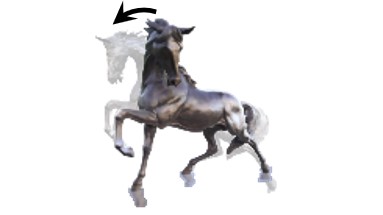} &
    \includegraphics[width=1.35\linewidth]{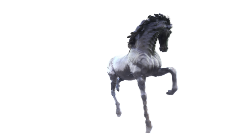} &
    \includegraphics[width=1.35\linewidth]{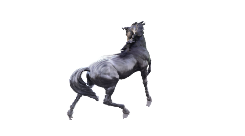} &
    \includegraphics[width=1.35\linewidth]{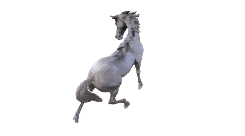}
    \\
    \vspace{-12mm} \rotatebox{90}{After} &
    \includegraphics[width=1.35\linewidth]{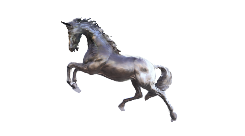} &
    \includegraphics[width=1.35\linewidth]{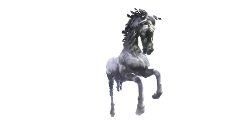} &
    \includegraphics[width=1.35\linewidth]{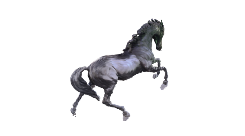} &
    \includegraphics[width=1.35\linewidth]{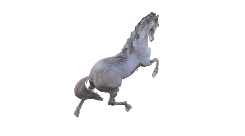}
    \\
    \vspace{-15mm} \rotatebox{90}{Before} &
    \includegraphics[width=1.05\linewidth]{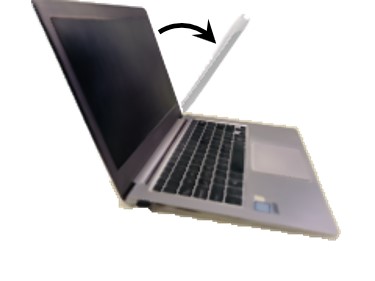} &
    \includegraphics[width=1.05\linewidth]{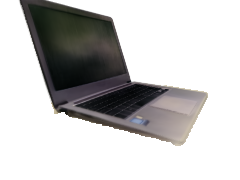} &
    \includegraphics[width=1.05\linewidth]{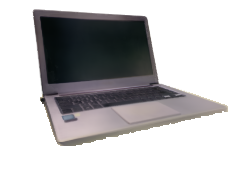} &
    \includegraphics[width=1.05\linewidth]{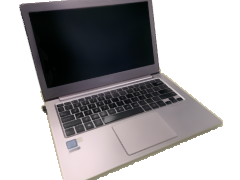}
    \\
    \vspace{-13mm} \rotatebox{90}{After} &
    \includegraphics[width=1.05\linewidth]{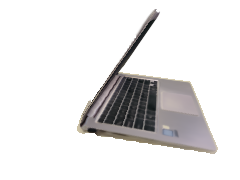} &
    \includegraphics[width=1.05\linewidth]{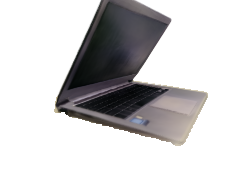} &
    \includegraphics[width=1.05\linewidth]{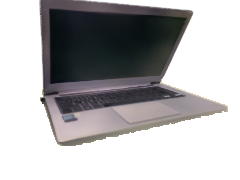} &
    \includegraphics[width=1.05\linewidth]{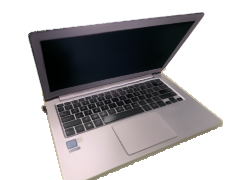}
    \\
    \vspace{-15mm} \rotatebox{90}{Before} &
    \includegraphics[width=1.3\linewidth]{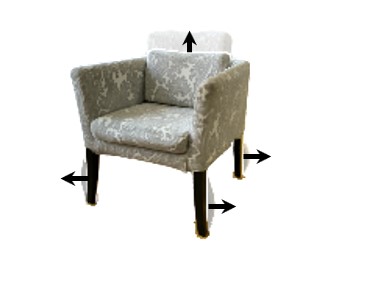} &
    \includegraphics[width=1.3\linewidth]{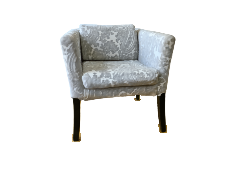} &
    \includegraphics[width=1.3\linewidth]{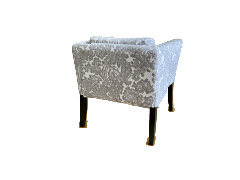} &
    \includegraphics[width=1.3\linewidth]{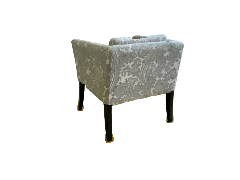}
    \\
    \specialrule{0em}{-3pt}{-3pt}
    \vspace{-15mm} \rotatebox{90}{After} &
    \includegraphics[width=1.3\linewidth]{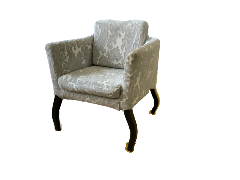} &
    \includegraphics[width=1.3\linewidth]{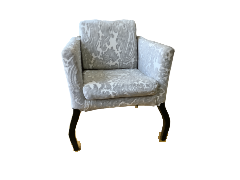} &
    \includegraphics[width=1.3\linewidth]{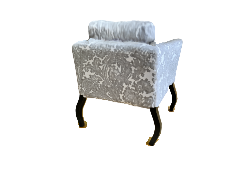} &
    \includegraphics[width=1.3\linewidth]{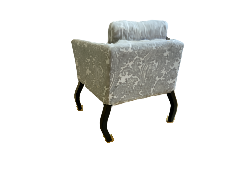}
    \end{tabular}
    \end{spacing}
    \vspace{-3mm}
    \caption{Results of our NeRF editing framework (row ``After'') compared with original NeRF results (row ``Before'') on multiple real captured data. Different columns show different views. We edit the static neural radiance fields and exhibit the deformed results under different views.}
    \vspace{-3mm}
    \label{fig:editing_results_real}
\end{figure}

\textbf{%
Shape editing results under different views.}
We first show NeRF editing results rendered from different views in Figs.~\ref{fig:editing_results}- \ref{fig:editing_results_real} for synthetic data and real captured objects. For comparison, we also show the results under the same views before editing. In Fig.~\ref{fig:editing_results}, the first set is a Lego bulldozer from the NeRF dataset. We put down its shovel and achieve the editing of complex synthetic data.
The second set is a synthetic chair from the NeRF dataset. We stretch the back and legs of the chair, which demonstrates that our method can edit the local parts of man-made objects. 
{In Fig.~\ref{fig:editing_results_giraffe}, we present the editing results on a giraffe soft toy captured by ourselves. It can be seen that users can edit the giraffe to different poses, as well as scale local areas, which demonstrates the usability of our method.}
{In Fig.~\ref{fig:editing_results_real}, we show four more sets of results from real scenes to illustrate that our method can be applied to different objects.} The wings of the toy dragon are deformed to make them spread out. This can further realize the animation of the dragon flapping its wings while viewing it from different directions.
We also show an example of a horse statue from the FVS dataset, 
where we can deform the horse's head and raise its hoof.
{On the example of a laptop, we can rotate its panel to be at different angles.}
{For the real captured chair, we bend the legs of the chair to present another design style, and at the same time enlarge the backrest, which make the chair more comfortable to sit on.}
These results show that our approach is able to deform static neural radiance fields according to the user's editing.
{In Fig.~\ref{fig:teaser}, we show an example of shape deformation for multiple objects in a scene. We first split the mesh of the horse statue from the scene, then copy it into multiple ones, place them in different locations, and deform them differently.}

\textbf{Deformation transfer results.}
In addition to 
{user-controlled shape deformation,}
we can also use deformation transfer methods to transfer the deformations from the existing deformation sequence to the real captured objects. This can achieve some interesting applications. For example, we can transfer the movements of a human face from a video clip to a head sculpture. We show some example results in supplementary material.

\subsection{Comparisons} \label{sec:compare}
As we are the first to perform general geometric {shape deformation} on NeRF, we propose {three} baseline methods for comparisons. For the first baseline, we adopt a naive way to build the correspondence between the extracted triangular mesh and continuous volume space. We no longer construct a tetrahedral mesh and use it as a proxy, but instead directly find the closest point of the sampled point on the extracted triangular mesh surface, and use the displacement of the closest point as the displacement of the sampled point. We denote the first one as ``Closest Point''. {The second baseline is similar to the first one. The difference is that we linearly interpolate the displacements of three nearest triangular mesh vertices, with the coefficients inversely related to distances, to obtain the displacement of the sampled point. We denote this one as ``3NN''. The last baseline} is mesh rendering. The extracted triangular mesh is with vertex color information, which can be directly rendered after 
{user-controlled shape deformation}
or deformation transfer. 

\begin{figure}[htb]
    \centering
    \setlength{\fboxrule}{0.5pt}
    \setlength{\fboxsep}{-0.01cm}
    \begin{spacing}{1}
    \setlength{\tabcolsep}{-0.1pt}
    \begin{tabular}{p{0.24\linewidth}<{\centering}p{0.24\linewidth}<{\centering}p{0.24\linewidth}<{\centering}p{0.24\linewidth}<{\centering}}

     GT & Closest Point & 3NN &  Ours  \\
    
    \includegraphics[width=1.05\linewidth]{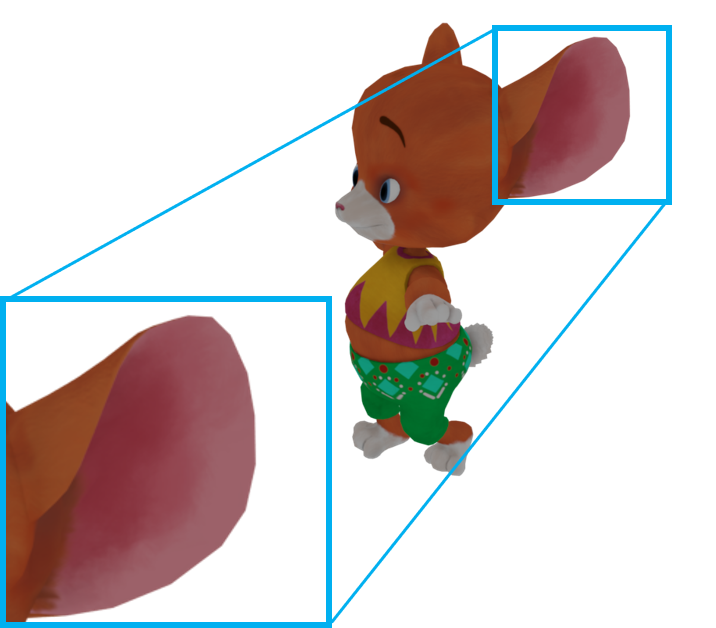} &
    \includegraphics[width=1.05\linewidth]{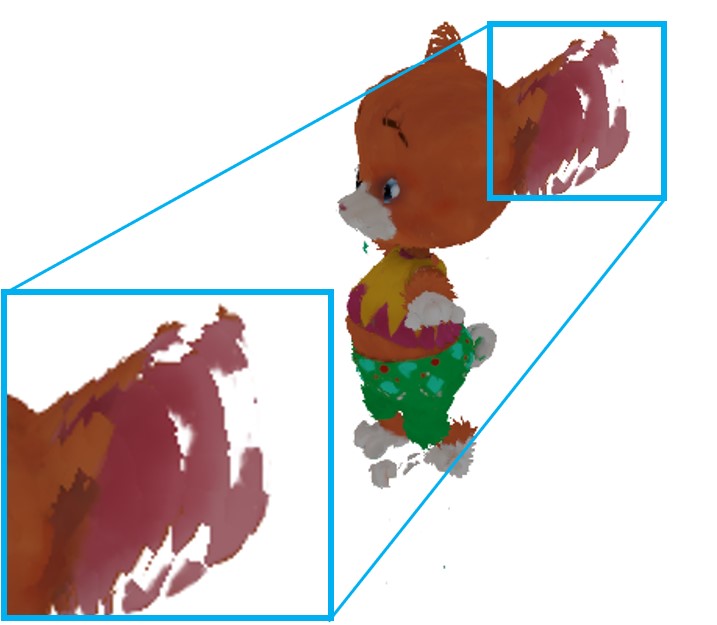} &
    \includegraphics[width=1.05\linewidth]{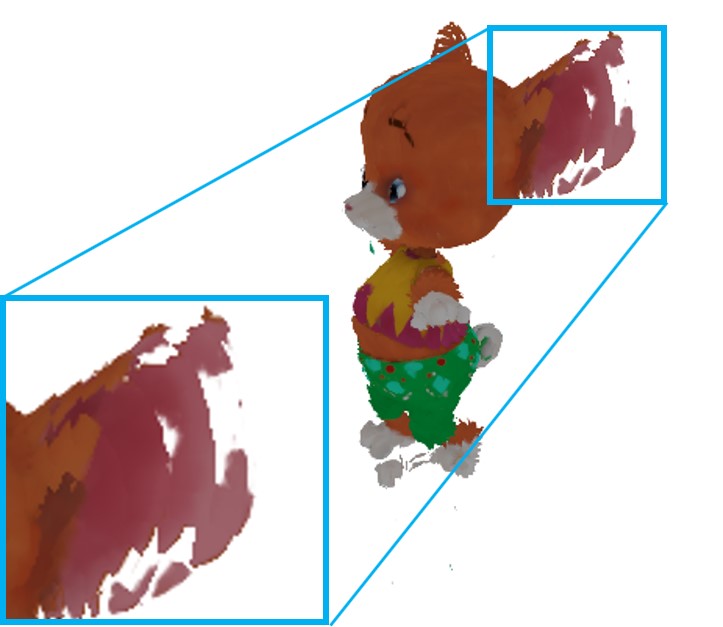} &
    \includegraphics[width=1.05\linewidth]{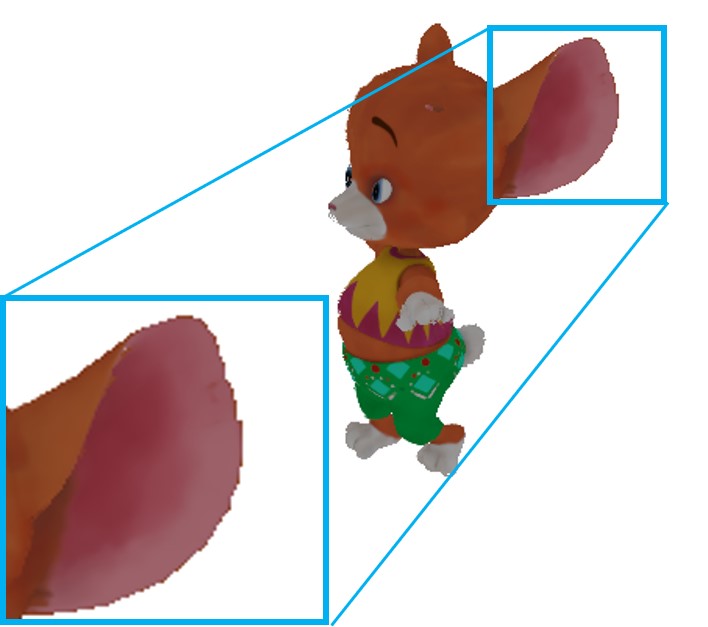}
    \\
    \specialrule{0em}{-2pt}{-2pt}
    \includegraphics[width=1.05\linewidth]{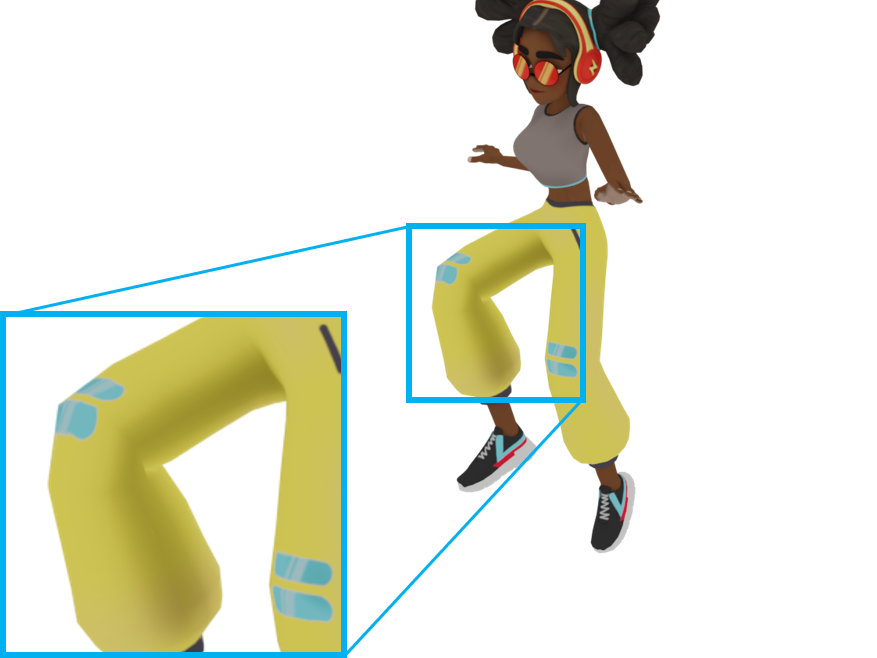} &
    \includegraphics[width=1.05\linewidth]{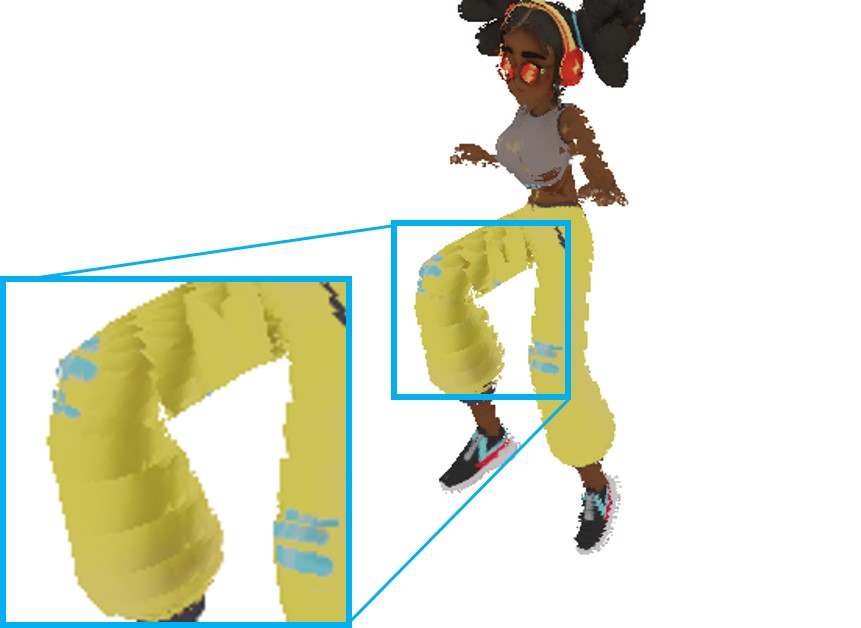} &
    \includegraphics[width=1.05\linewidth]{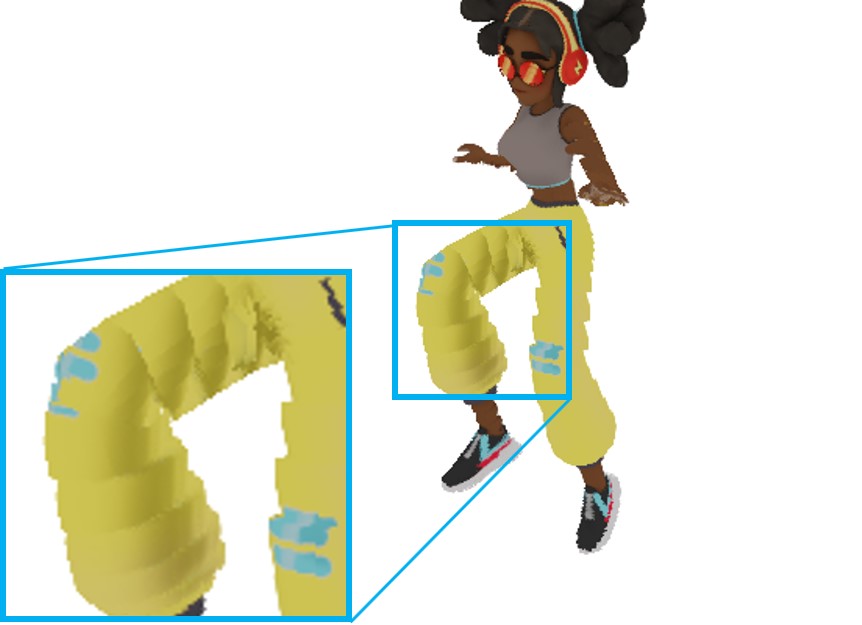} &
    \includegraphics[width=1.05\linewidth]{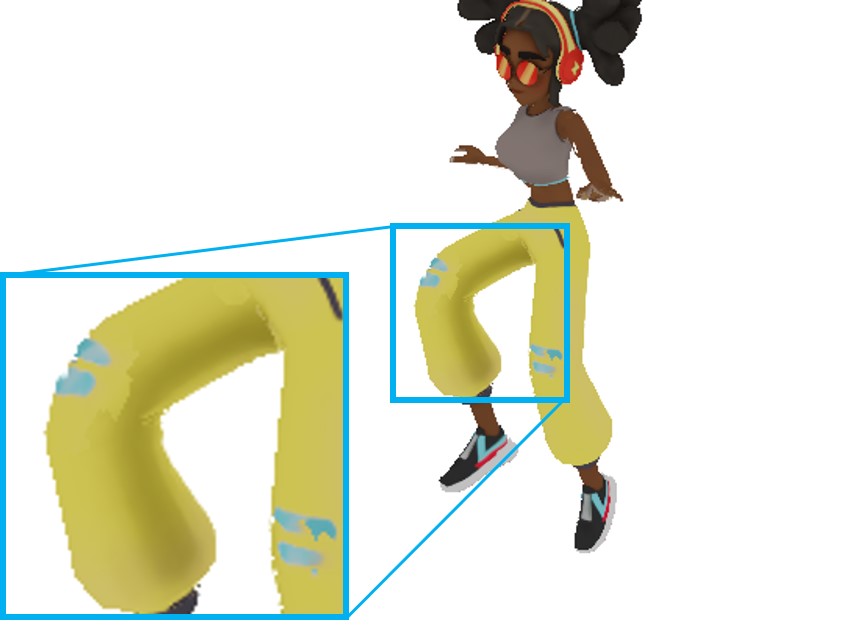}
    \\
    \specialrule{0em}{-1pt}{-1pt}
    \includegraphics[width=1.05\linewidth]{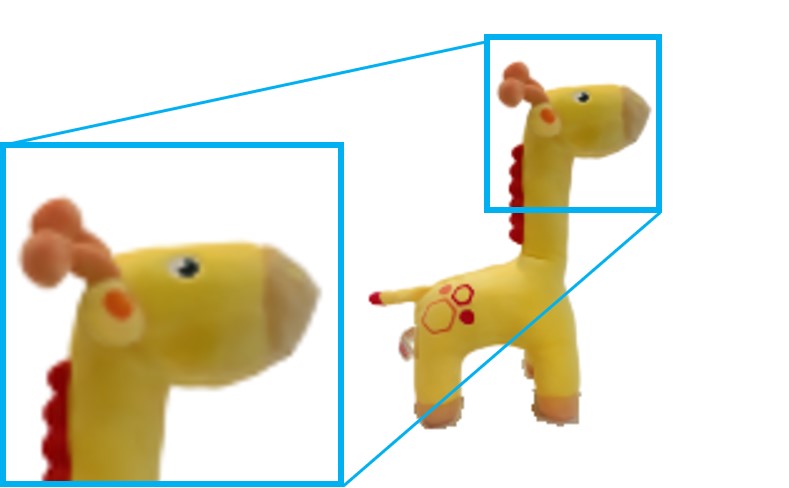} &
    \includegraphics[width=1.05\linewidth]{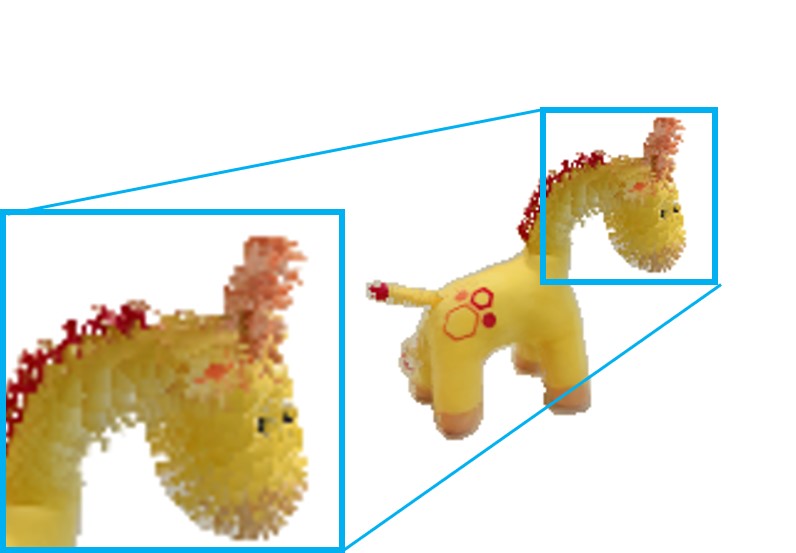} &
    \includegraphics[width=1.05\linewidth]{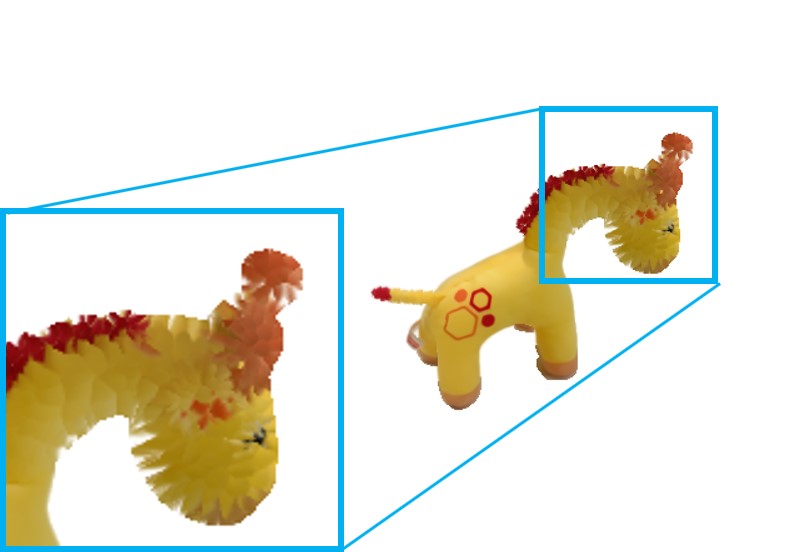} &
    \includegraphics[width=1.05\linewidth]{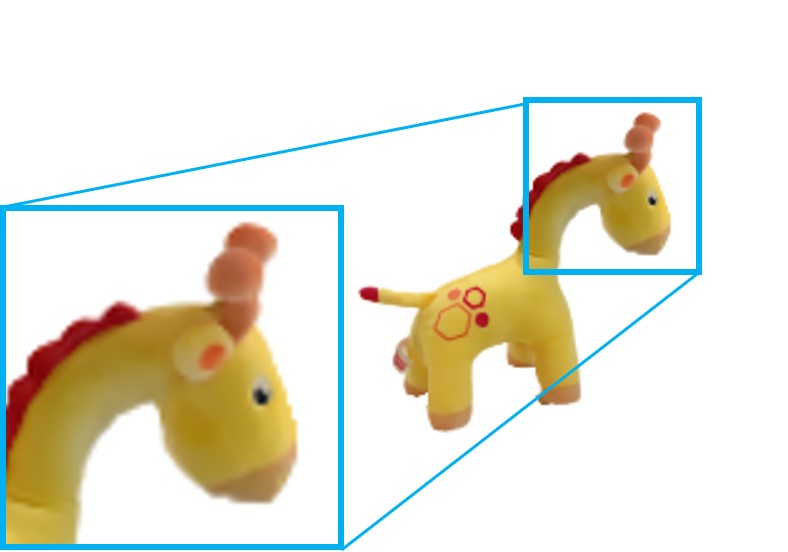}

    \end{tabular}
    \end{spacing}
    \vspace{-3mm}
    \caption{{Visual comparisons with the ``Closest Point'' and ``3NN'' baselines. The ``Closest Point'' and ``3NN'' baselines may cause discontinuities, so the resulting rendering results have obvious artifacts. Note that the last row is a real captured toy giraffe, so ground truth does not exist and we instead visualize the NeRF rendering results before deformation for reference.}}
    \vspace{-4mm}
    \label{fig:comparison}
\end{figure}

{We compare our method with the ``Closest Point'' and ``3NN'' baselines} on the synthetic data mixamo which has ground truth edited results. The visual comparison results are shown in Fig.~\ref{fig:comparison}, and the quantitative comparison is shown in Table~\ref{table:comparison}. 
We also show a visual comparison on a real captured scene in the last row of Fig.~\ref{fig:comparison}. Due to the absence of ground truth, we visualize the NeRF rendering result before deformation and corresponding deformation results.
It can be seen that {the ``Closest Point'' and ``3NN'' baselines} may cause discontinuities, so the rendering results have obvious artifacts, while our method adopts two-step deformation transfer, and the results are more visually satisfactory and has quantitative advantages. 

\begin{table}[t]
\begin{center}
\resizebox{\linewidth}{!}{
\begin{tabular}{lcccc}
\hline
Method        & SSIM $\uparrow$  &  LPIPS $\downarrow$  &  PSNR $\uparrow$  & FID (real) $\downarrow$\\ \hline
Closest Point       & 0.928          & 0.055           & 22.38         & 300.8\\
3NN        & 0.941          & 0.042           & 24.30   
      & 291.7 \\
Ours            &   \textbf{0.975}        & \textbf{0.024}  & \textbf{29.62} & \textbf{253.7}  \\ \hline
\end{tabular}
}
\end{center}
\vspace{-5mm}
\caption{{Quantitative comparison with the ``Closest Point'' and ``3NN'' baselines. It can be seen that our framework achieves better results. Note that the first three metrics are calculated on mixamo synthetic data, while the last FID is calculated on real data.}}
\vspace{-6mm}
\label{table:comparison}
\end{table}

\begin{figure}[htb]
    \centering
    \setlength{\fboxrule}{0.5pt}
    \setlength{\fboxsep}{-0.01cm}
    \begin{spacing}{1}
    \setlength{\tabcolsep}{-0.2pt}
    \begin{tabular}{p{0.25\linewidth}<{\centering}p{0.25\linewidth}<{\centering}p{0.25\linewidth}<{\centering}p{0.25\linewidth}<{\centering}}

     \small{Mesh Render} & \small{NeRF} & \small{Edited Mesh} &  \small{Our Editing}  \\

    \includegraphics[width=1.15\linewidth]{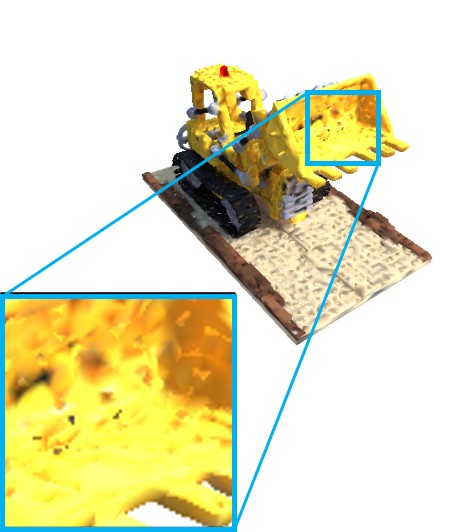} &
    \includegraphics[width=1.15\linewidth]{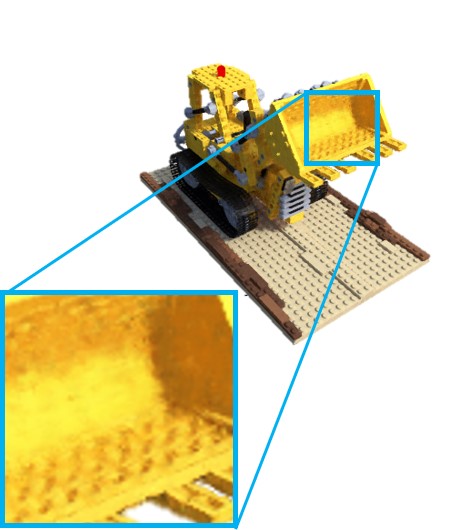} & 
    \includegraphics[width=1.15\linewidth]{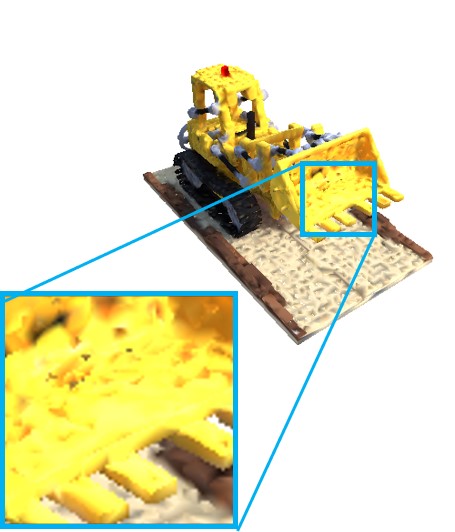} &
    \includegraphics[width=1.15\linewidth]{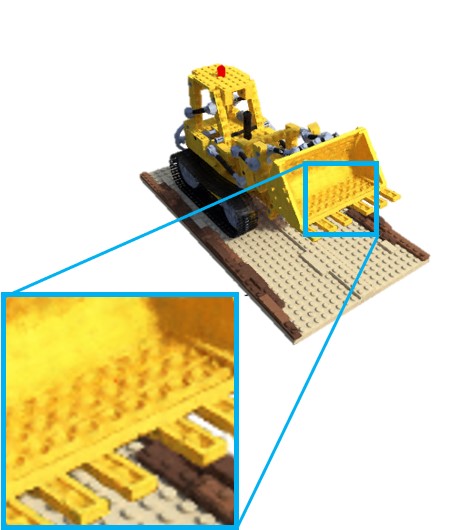}

    \end{tabular}
    \end{spacing}
    \vspace{-3mm}
    \caption{Comparisons with mesh rendering on the synthetic data. Mesh rendering has obvious artifacts when the mesh quality is not good, while this does not affect our editing and image synthesis.}
    \vspace{-6mm}
    \label{fig:comparison_sr}
\end{figure}

Then we compare our method with mesh rendering baseline on the Lego data from NeRF. It should be noted that although our method uses an explicit triangular mesh representation for interactive editing, it has a certain degree of error tolerance in terms of the mesh reconstruction, and the reconstructed triangular mesh does not need to be perfect. This is because the mesh is only used as an intermediate representation and our final images are still obtained through volume rendering. The direct mesh rendering requires a high quality mesh, and all artifacts on the mesh will appear in the rendered images. As shown in Fig.~\ref{fig:comparison_sr}, the reconstructed mesh in the lego is not of good quality, and the result of mesh rendering is not ideal, while our method can still perform editing, and with the help of volume rendering, desired results can still be obtained.

\subsection{Ablation Study}
We conduct ablation studies on the synthetic data with respect to the novel view synthesis results after editing. First, as we introduce a tetrahedral mesh in our method as a proxy between the triangular mesh and the continuous volume, we compare the results of editing on the triangular mesh and editing on the tetrahedral mesh, and verify the necessity of editing on the triangular mesh and transferring deformation by our method. Second, in order to evaluate the influence of the reconstructed triangular mesh on our results, we compare the results of the triangular mesh extracted by the original NeRF and that extracted by NeuS which improves the quality of reconstruction. Tables~\ref{table:edit_mesh} and \ref{table:mesh_quality} summarize the quantitative results of the ablation studies.

\begin{figure}[htb]
    \centering
    \setlength{\fboxrule}{0.5pt}
    \setlength{\fboxsep}{-0.01cm}
    \begin{spacing}{1}
    \begin{tabular}{p{0.3\linewidth}<{\centering}p{0.3\linewidth}<{\centering}p{0.3\linewidth}<{\centering}}

     GT & Tetrahedral &  Triangular  \\
    
    \includegraphics[width=1.05\linewidth]{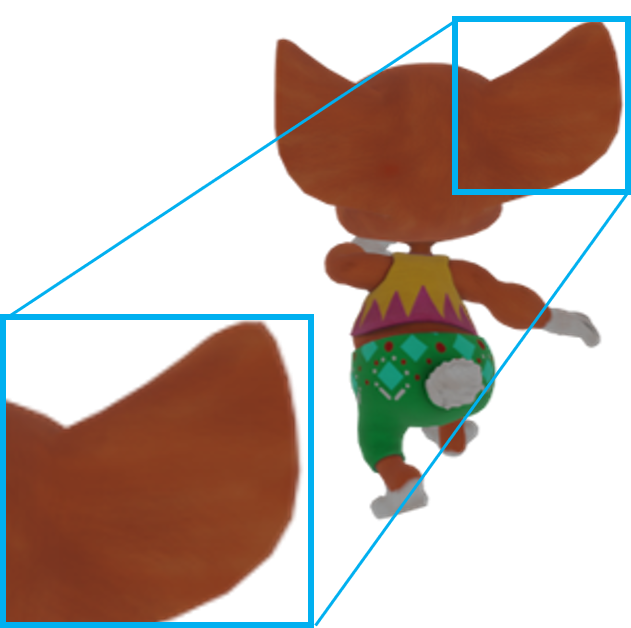} &
    \includegraphics[width=1.05\linewidth]{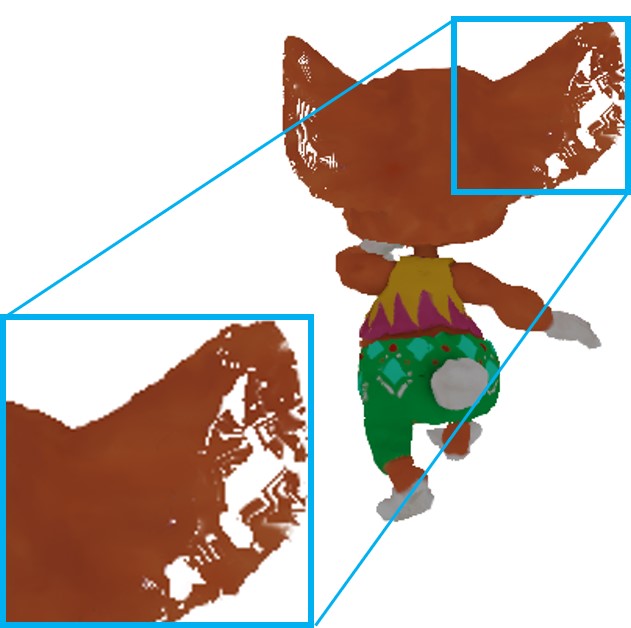} &
    \includegraphics[width=1.05\linewidth]{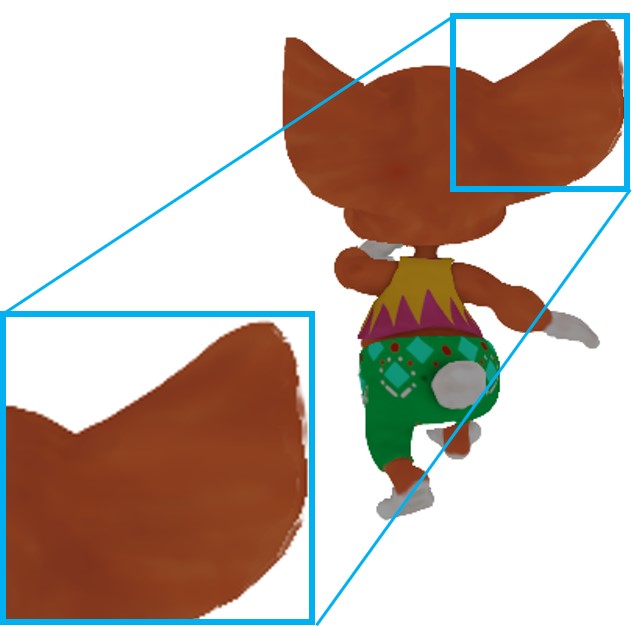}

    \end{tabular}
    \end{spacing}
    \vspace{-3mm}
    \caption{Ablation study of editing on the tetrahedral mesh or triangular mesh. It can be seen that editing on the tetrahedral mesh will bring in artifacts in rendered results.}
    \vspace{-6mm}
    \label{fig:edit_mesh}
\end{figure}

\textbf{Necessity of edit on triangular mesh.}
Table~\ref{table:edit_mesh} shows the quantitative comparisons between editing on the tetrahedral mesh and triangular mesh, which indicates that editing on triangular mesh performs better. The qualitative results are presented in Fig.~\ref{fig:edit_mesh}. The results of editing on tetrahedral mesh have obvious artifacts due to poor tetrahedral mesh quality. 

\begin{figure}[htb]
    \centering
    \setlength{\fboxrule}{0.5pt}
    \setlength{\fboxsep}{-0.01cm}
    \begin{spacing}{1}
    \setlength{\tabcolsep}{-1pt}
    \begin{tabular}{p{0.3\linewidth}<{\centering}p{0.3\linewidth}<{\centering}p{0.3\linewidth}<{\centering}}

     GT & NeRF &  NeuS  \\
    
    \includegraphics[width=1.05\linewidth]{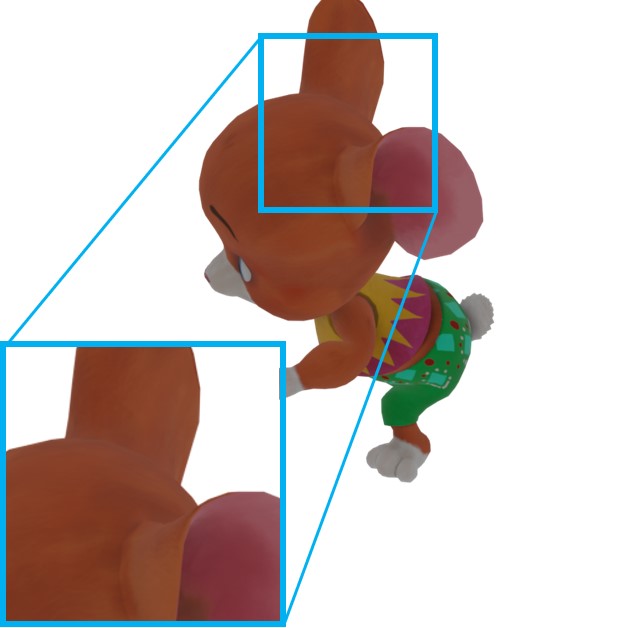} &
    \includegraphics[width=1.05\linewidth]{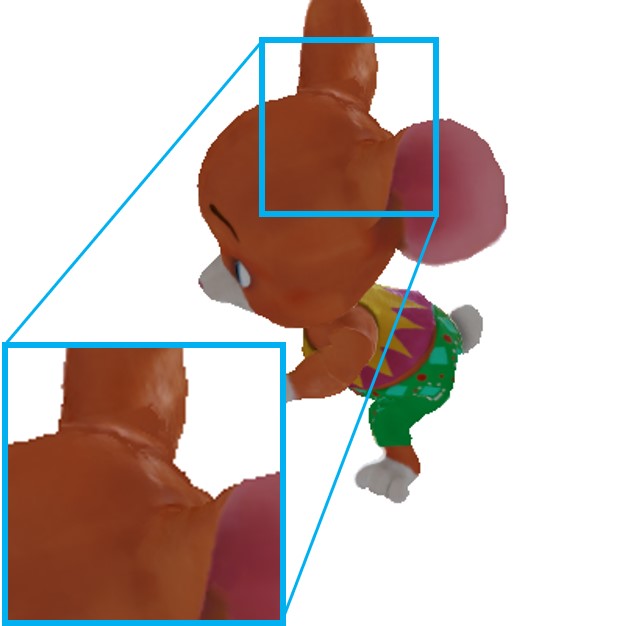} &
    \includegraphics[width=1.05\linewidth]{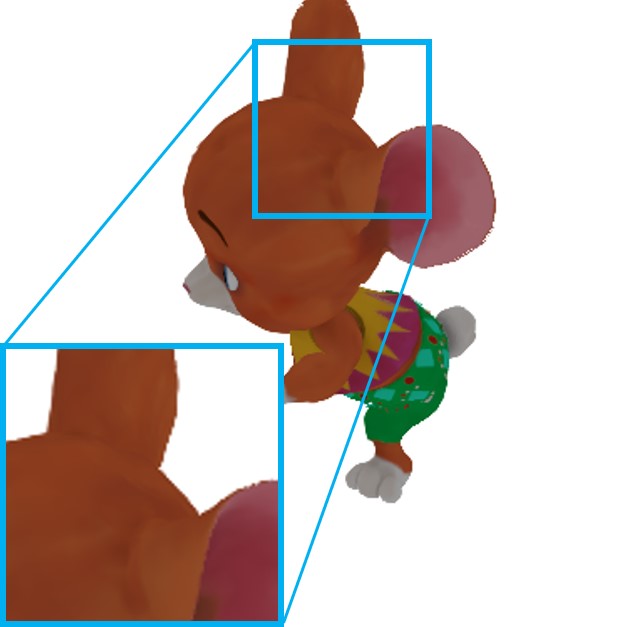}
    \\
    \specialrule{0em}{-2pt}{-2pt}
    \includegraphics[width=1.05\linewidth]{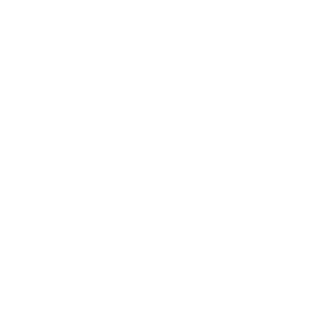} &
    \includegraphics[width=1.05\linewidth]{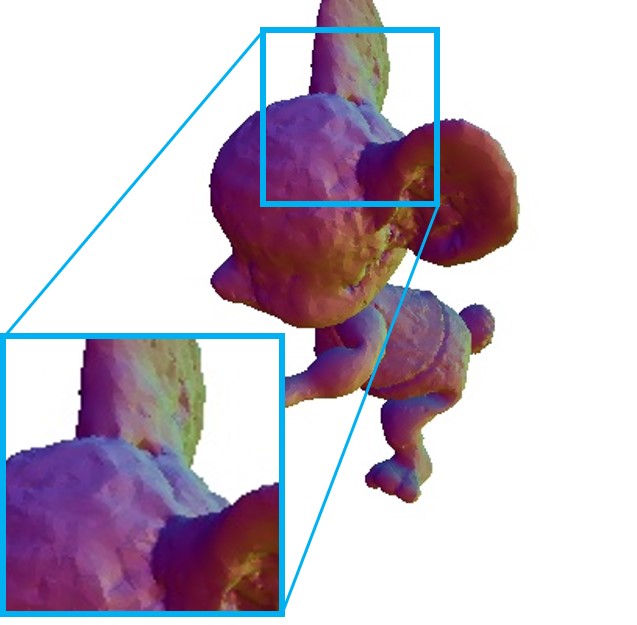} &
    \includegraphics[width=1.05\linewidth]{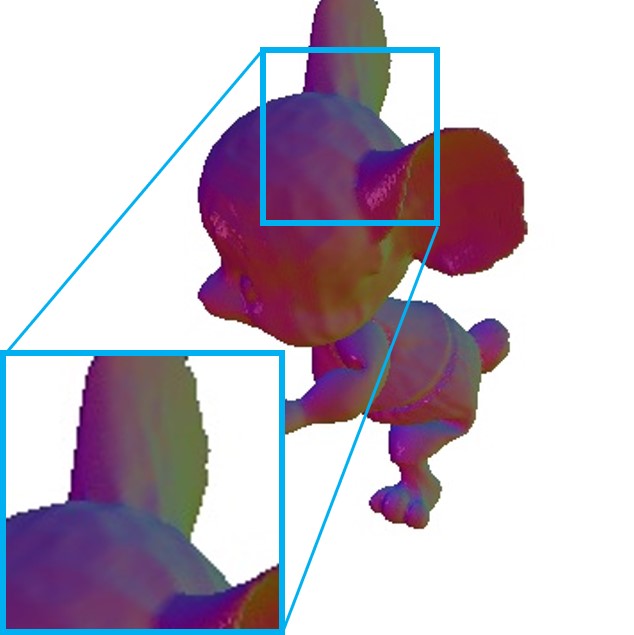}
    
    \end{tabular}
    \end{spacing}
    \vspace{-3mm}
    \caption{Ablation study of mesh quality. The reconstructed mesh from NeRF is worse than that of NeuS, leading to some artifacts in the rendered results. We visualize the rendered results (first row) and the mesh colored with vertex normals (second row).}
    \vspace{-3mm}
    \label{fig:mesh_quality}
\end{figure}

\textbf{Impact of mesh quality.}
Table~\ref{table:mesh_quality} evaluates the influence of the reconstructed mesh quality to our method. It can be seen that the result of using the mesh from NeuS is better than that of NeRF, but the difference is small. The visual comparisons are shown in Fig.~\ref{fig:mesh_quality}, where the results of using the mesh from NeRF have some artifacts in detail, but the overall result {is not bad}. This illustrates that mesh quality has little effect on our results.

\begin{table}[t]
\begin{center}
\begin{tabular}{lccc}
\hline
Method                   & SSIM $\uparrow$ & LPIPS $\downarrow$ & PSNR $\uparrow$ \\
\hline       
Edit on tetrahedral mesh & 0.934              & 0.049          & 24.37 \\
Edit on triangular mesh  & \textbf{0.975}     & \textbf{0.024} & \textbf{29.62}  \\
\hline
\end{tabular}
\end{center}
\vspace{-3mm}
\caption{Evaluation on the necessity of editing on the triangular mesh. Editing on the triangular mesh leads to better results than directly editing on the tetrahedral mesh.}
\vspace{-6mm}
\label{table:edit_mesh}
\end{table}

\begin{table}[h]
\begin{center}
\begin{tabular}{lccc}
\hline
Method  & SSIM $\uparrow$ & LPIPS $\downarrow$ & PSNR $\uparrow$ \\
\hline       
NeRF  & 0.969              & 0.027        & 28.95  \\ 
NeuS  & \textbf{0.975}     & \textbf{0.024}        & \textbf{29.62}   \\
\hline
\end{tabular}
\end{center}
\vspace{-6mm}
\caption{Evaluation on the impact of the extracted mesh quality. The reconstructed mesh from NeuS is better than that from NeRF, leading to better editing results.}
\vspace{-6mm}
\label{table:mesh_quality}
\end{table}

\subsection{Limitations}
\begin{figure}[htb]
    \centering
    \setlength{\fboxrule}{0.5pt}
    \setlength{\fboxsep}{-0.01cm}
    \begin{spacing}{1}
    \setlength{\tabcolsep}{-1pt}
    \begin{tabular}{p{0.3\linewidth}<{\centering}p{0.3\linewidth}<{\centering}p{0.3\linewidth}<{\centering}}

     Training1 & Training2 &  Deformed  \\
    
    \includegraphics[width=1.05\linewidth]{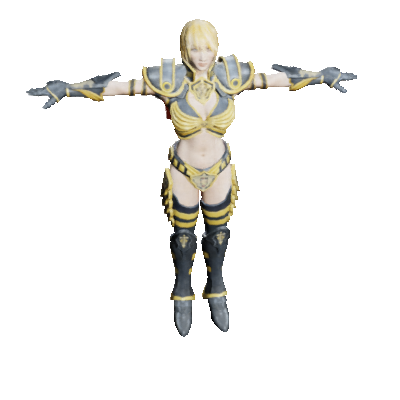} &
    \includegraphics[width=1.05\linewidth]{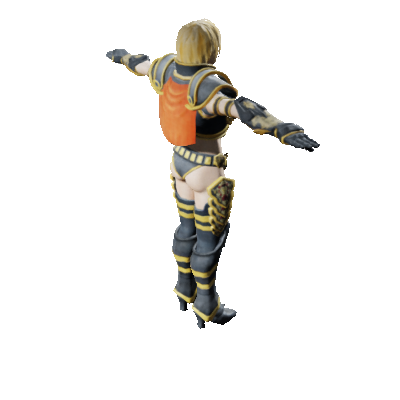} &
    \includegraphics[width=1.05\linewidth]{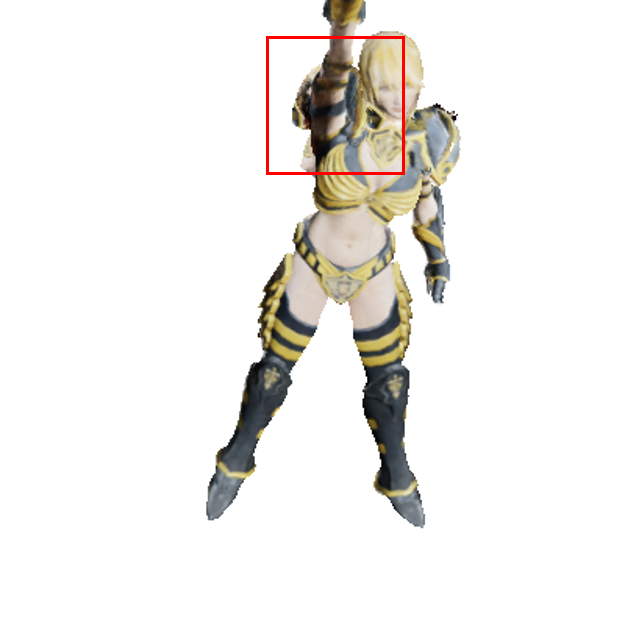} \\
    \specialrule{0em}{-3pt}{-3pt}
    
    \end{tabular}
    \end{spacing}
    \vspace{-6mm}
    \caption{Failure case. Our approach does not edit the appearance along with geometry deformation. In this example, since the underarm is occluded from the light in the T-pose during training, it will always be gloomy even when the woman raises her arm, which is implausible.}
    \vspace{-6mm}
    \label{fig:limitation}
\end{figure}

Our method is the first step for {geometric shape deformation} on NeRFs and still has several limitations. First of all, the biggest limitation is that we cannot modify the color and also the light and shadow based on the editing results. If an object part that is in the shadow during capturing is deformed to face the light, its color will still be dim instead of bright, as shown in Fig.~\ref{fig:limitation}. This could be dealt with by incorporating some recent NeRF-based relighting work~\cite{boss2021nerd,zhang2021nerfactor} to achieve correct color rendering by decoupling lighting. Second, our method cannot support real-time editing by users. The user can only select a viewing angle for image synthesis after editing the mesh representation. At present, the main time bottleneck is still in the rendering part of NeRF. Recently, there are some works on the acceleration of NeRF rendering~\cite{garbin2021fastnerf,yu2021plenoctrees,hedman2021baking}. The combination with these methods will help with real-time rendering of interactive editing results.

\section{Conclusion}
In this paper, we propose the first method to support 
{user-controlled shape deformation to}
the geometry of neural radiance field network. By establishing a correspondence between the explicit mesh representation and the implicit volume representation, our method can use the well-developed triangular mesh deformation method to deform the implicit representation. With the novel view synthesis capability of NeRF, users can visualize the editing results from arbitrary views. Our method is suitable for general real scenes which can edit scene objects including human bodies, animals, man-made models, etc. Compared with the previous editing methods for NeRF, our method has a higher degree of freedom and can support the editing of details. In the future, we will further explore the combination of relighting methods. After editing the scene geometry, the corresponding colors can be modified to make the light and shadow in the rendering results more natural. {In future work, we will implement our proposed approach in Jittor~\cite{hu2020jittor}, which is a fully just-in-time (JIT) compiled deep learning framework.} 
\vspace{-6mm}

\section*{Acknowledgement}
\noindent This work was supported by the Beijing Municipal Natural Science Foundation for Distinguished Young Scholars (No. JQ21013), the National Natural Science Foundation of China (No. 62061136007 and No. 61872440), Royal Society Newton Advanced Fellowship (No. NAF\verb|\|R2\verb|\|192151), the Alibaba Innovative Research (AIR) Program and the Youth Innovation Promotion Association CAS.

{\small
\bibliographystyle{ieee_fullname}
\bibliography{egbib}
}

\end{document}